\documentclass{emulateapj}

\begin{document}

\shortauthors{Luhman}
\shorttitle{Brown Dwarfs in Taurus}

\title{The Spatial Distribution of Brown Dwarfs in Taurus}

\author{K. L. Luhman}
\affil{Department of Astronomy and Astrophysics,
The Pennsylvania State University, University Park, PA 16802;
kluhman@astro.psu.edu.}

\begin{abstract}

By combining photometry from the 2MASS Point Source Catalog and the 
USNO-B1.0 Catalog with optical and infrared spectroscopy, I have 
performed a search for young brown dwarfs in an area of 225~deg$^2$ 
encompassing all of the Taurus star-forming region ($\tau\sim1$~Myr).
From this work, I have discovered 22 new members of Taurus, five 
of which were independently found by Guieu and coworkers.
Sixteen of these new members have spectral types later than M6 and thus
are likely to be brown dwarfs according to the theoretical evolutionary 
models of Chabrier and Baraffe.
After adding these new members to the previously known members of Taurus,
I have compared the spatial distributions of stars and brown dwarfs across
the entire region.
I find no statistically significant difference between these two distributions. 
Taurus does not contain the large, extended population of brown dwarfs that 
has been predicted by some embryo ejection models for the formation of brown 
dwarfs. However, these results are consistent with other ejection 
models, as well as models in which stars and brown dwarfs share a common 
formation mechanism.

\end{abstract}

\keywords{infrared: stars --- stars: evolution --- stars: formation --- stars:
low-mass, brown dwarfs --- stars: luminosity function, mass function ---
stars: pre-main sequence}

\section{Introduction}
\label{sec:intro}

Measuring the spatial distribution of a population of newly-formed stars 
can offer insight into the star formation process.
Because of its youth and low stellar density, the stellar content
within the Taurus-Auriga molecular cloud probably has undergone the least 
dynamical evolution of any nearby star-forming region, and therefore
is the best available site for measuring a primordial spatial distribution 
of young stars.
Several studies over the last decade have taken advantage of this fact
to address various aspects of star formation. For instance, 
measurements of the distribution of stars in Taurus have revealed effects
of both binary formation and the Jeans condition on clustering properties
\citep{gom93,lar95,sim97,nak98,bat98}
while a comparison of the distributions of stars and molecular material in
Taurus has constrained the velocity dispersion with which
the stars are born and the nature of cloud fragmentation \citep{har02}. 

Measuring the spatial distribution of Taurus members as a function of mass
is another potentially rewarding experiment. 
Several studies have suggested that brown dwarfs might form as protostellar 
sources whose accretion is prematurely halted by ejection from multiple 
systems \citep{rc01,bos01,bat02,del03,umb05}.
Some of these ejection models have predicted that newborn brown dwarfs 
could have higher velocity dispersions than their stellar counterparts 
\citep{rc01,kb03}. If so, brown dwarfs would be more widely distributed 
than stars in star-forming regions. 
For instance, \citet{kb03} favored an ejection model with a one-dimensional 
velocity dispersion of $\sim2$~km~s$^{-1}$, which corresponds to an angular
distance of $0\fdg8$ for an object traveling for 1~Myr at the distance of 
Taurus. Meanwhile, other models of ejection predict that stars and brown dwarfs 
should have similar spatial and velocity distributions \citep{bat03}.

Over time, surveys for substellar members of Taurus have encompassed larger 
areas surrounding the stellar aggregates.  
Some of these data have exhibited no statistically significant differences in 
the spatial distributions of the high- and low-mass members of Taurus
\citep{bri02,luh04tau} while other data have suggested a possible difference
\citep{gui06}.
To search for an extended population of brown dwarfs beyond the previous 
survey fields and to measure the distribution of brown dwarfs in Taurus on
the largest size scales, I have performed a survey for brown dwarfs across an
area of 225~deg$^2$ containing the entire Taurus star-forming region. 
In this paper, I describe the selection of candidate substellar
members of Taurus (\S~\ref{sec:cand}) and the spectroscopy and classification
of these candidates (\S~\ref{sec:spec}), evaluate the completeness of this
survey and the recent one by \citet{gui06} (\S~\ref{sec:complete}), 
and use the resulting updated census of Taurus to perform a definitive
comparison of the distributions of stars and brown dwarfs in this region
(\S~\ref{sec:spatial}).

\section{Selection of Brown Dwarf Candidates}
\label{sec:cand}

\begin{figure}
\plotone{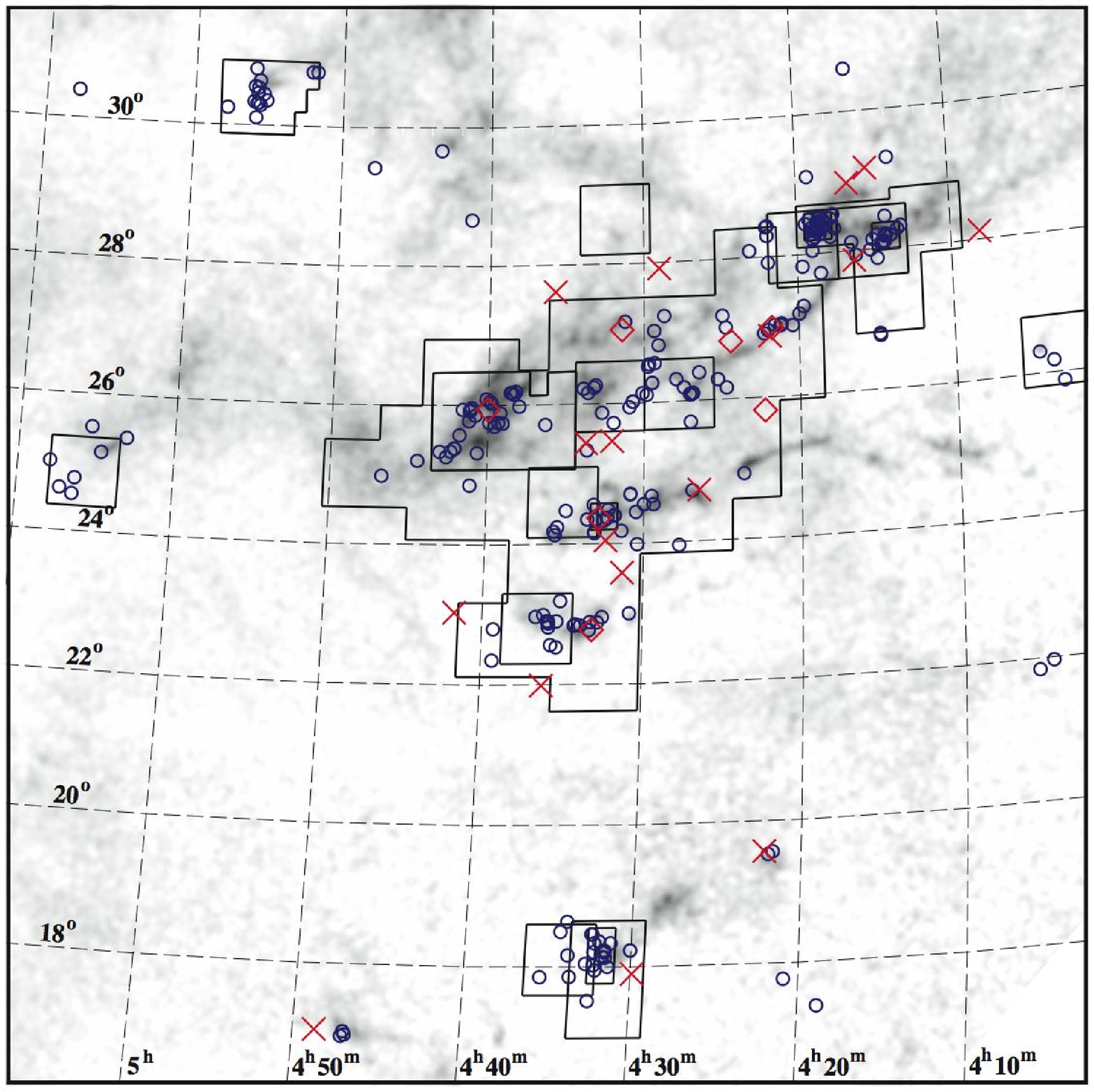}
%\plotone{../psfiles/map1.ps}
\caption{
Spatial distribution of previously known members of the Taurus
star-forming region ({\it circles}), members from \citet{gui06} that
have been independently discovered in this work ({\it diamonds}), and
the additional new members from this work ({\it crosses}) shown with a map 
of extinction \citep[{\it grayscale},][]{dob05}.
The regions previously surveyed for brown dwarfs are indicated
\citep[{\it lines},][]{bri98,bri02,luh00tau,luh04tau,luh03tau,gui06}.
In this work, the entire area shown ($15\arcdeg\times15\arcdeg$) has 
been searched for brown dwarfs.
}
\label{fig:map1}
\end{figure}

\begin{figure}
\plotone{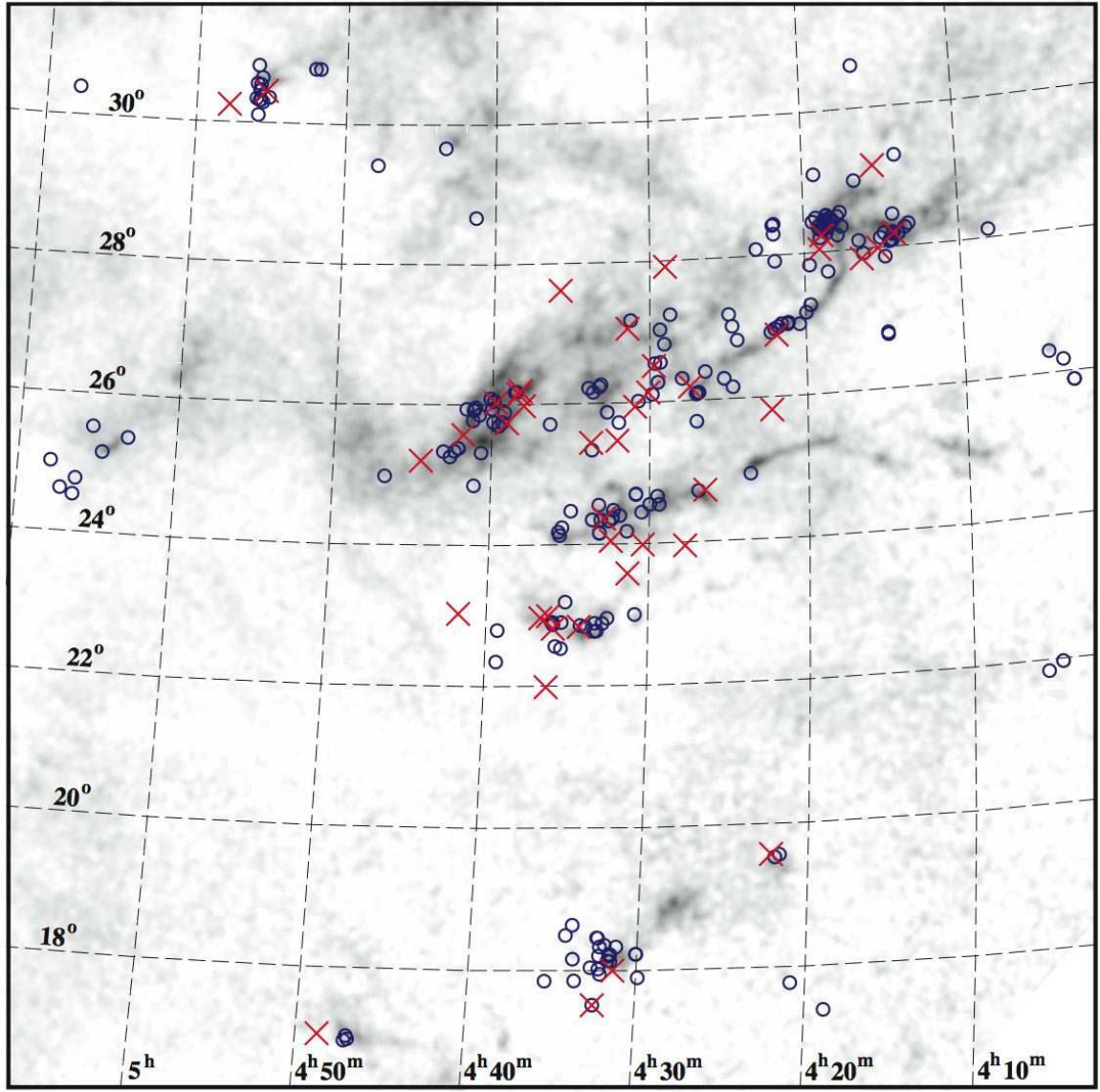}
%\plotone{../psfiles/map2.ps}
\caption{
Spatial distributions of stars ($\leq$M6, {\it circles}) 
and brown dwarfs ($>$M6, {\it crosses}) 
in the Taurus star-forming region shown with a map
of extinction \citep[{\it grayscale},][]{dob05}.
}
\label{fig:map2}
\end{figure}

Previous surveys for low-mass stars and brown dwarfs in the Taurus
star-forming region have identified candidates through color-magnitude 
and color-color diagrams constructed from optical and near-infrared (IR)
photometry 
\citep[][hereafter G06]{luh00tau,luh04tau,mar01,bri98,bri02,luh03tau,gui06}.
Those optical data were obtained through dedicated imaging with wide-field
cameras of areas ranging from $\sim1$ to 28~deg$^{2}$. 
IR measurements for these optical fields were then taken from 
the Two-Micron All-Sky Survey \citep[2MASS,][]{skr06}.
2MASS is also the source of IR photometry for the survey in this work.
However, as a substitute for optical CCD photometry, I use the photometric 
measurements designated as the second epoch near-IR magnitude (or photographic 
$I$) in the USNO-B1.0 Catalog \citep{mon03}, which is referred to as $I2$
in this work. 
By employing only all-sky catalogs, one can survey a field of any area and
position. Therefore, I consider a field that is large enough to encompass
all of the Taurus star-forming region, which I choose to center at
$\alpha=4^{\rm h}36^{\rm m}00^{\rm s}$, $\delta=24\arcdeg00\arcmin00\arcsec$
(J2000) with dimensions of $15\arcdeg\times15\arcdeg$.
The boundaries of this survey area correspond approximately to the boundaries
of the maps in Figs.~\ref{fig:map1} and \ref{fig:map2}.

\begin{figure}
\plotone{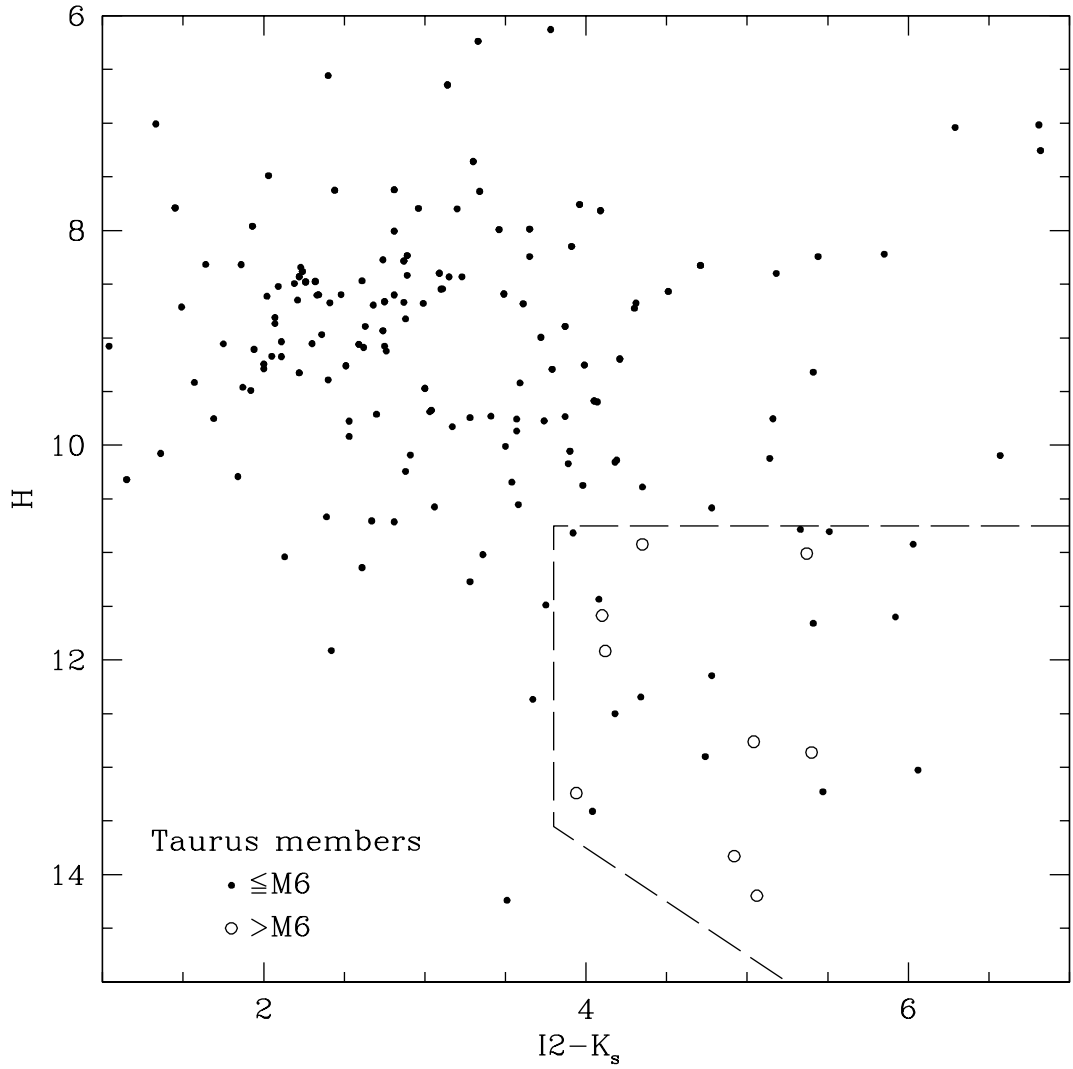}
\caption{
Color-magnitude diagram for known members of the Taurus star-forming region
constructed from photometry in the 2MASS ($H$, $K_s$) and USNO-B1.0 ($I2$)
catalogs. The known substellar members of Taurus 
({\it circles}, $>$M6) have been used to 
define ranges of magnitudes and colors ({\it dashed line}) to act as criteria
for identifying new candidate brown dwarfs in the $15\arcdeg\times15\arcdeg$
area shown in Figs.~\ref{fig:map1} and \ref{fig:map2}.
}
\label{fig:ik}
\end{figure}

\begin{figure}
\plotone{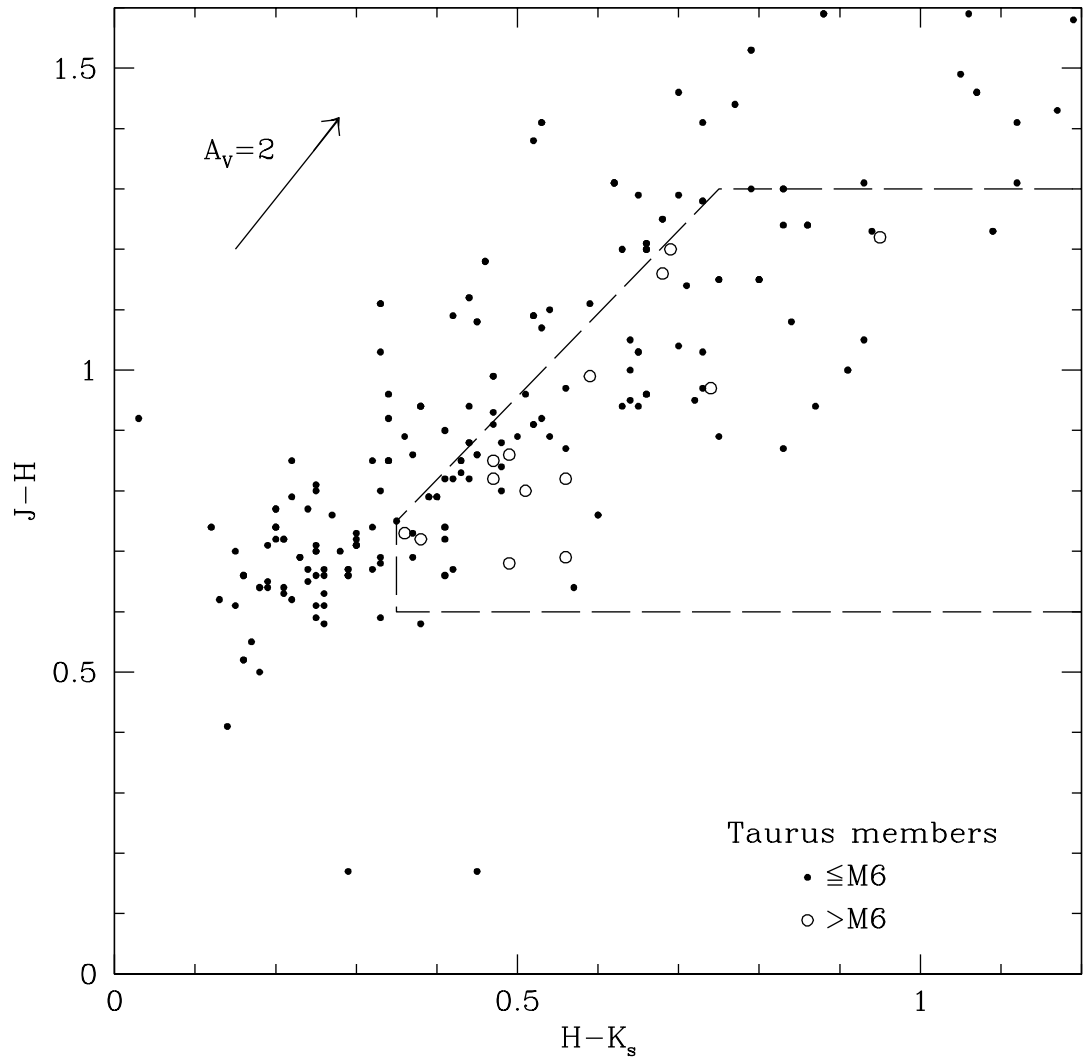}
\caption{
Same as Figure~\ref{fig:ik} for $J-H$ versus $H-K_s$.
}
\label{fig:jhhk}
\end{figure}

To develop criteria for identifying candidate substellar members of Taurus 
with data from the USNO and 2MASS catalogs, I used the 14 known members 
with spectral types later than M6\footnote{The hydrogen burning mass limit
at ages of 0.5-3~Myr corresponds to a spectral type of $\sim$M6.25
according to the models of \citet{bar98} and \citet{cha00} and the
temperature scale of \citet{luh03b}.} from \citet{luh03tau} and prior studies
that are resolved by 2MASS (excludes GG~Tau~Bb).
Nine of these late-type members have measurements in $I2$, while the 
remaining five sources are below the detection limit. All of the known 
late-type members of Taurus have measurements in the 2MASS catalog.
The sources detected at $I2$ are shown in a diagram of $H$ versus $I2-K_s$ in
Figure~\ref{fig:ik}. 
The ranges of colors and magnitudes that encompass these objects can
be defined as $I2-K_s>3.8$, $I2-K_s>H-9.75$, and $H\geq10.75$.
I have performed a similar exercise with $J-H$ versus $H-K_s$ in 
Figure~\ref{fig:jhhk}, where the 14 members later than M6 exhibit colors of
$0.6\leq J-H\leq1.3$, $H-K_s\geq0.35$, and $H-K_s\geq0.73(J-H-0.75)+0.35$.
In addition, the objects with $H>13$ have $H-K_s>0.45$, 
and 12 of the 14 late-type members have photometric uncertainties less 
than 0.05~mag in all three bands of the 2MASS data.

To search for new late-type members of Taurus that are in the same range
of masses and extinctions as the previously known ones, I applied all 
of the above criteria to the 2MASS Point 
Source Catalog and the USNO-B1.0 Catalog for the 225~deg$^2$
field encompassing Taurus.
For objects not detected in $I2$, only the criteria using the 2MASS
photometry were applied. This process produced 112 candidates.
I rejected 21 candidates that appeared to be field stars based on 
color-magnitude diagrams from \citet{luh00tau}, \citet{bri02}, and 
\citet{luh04tau}, 
six candidates identified and presented in the concurrent survey by 
\citet{luh04tau}, and two candidates for which $I2$ was unavailable
because of close proximity to a brighter star rather than a faint flux
in that band.
The remaining 83 candidates comprised the final sample.
\citet{luh04tau} presented spectroscopy of 15 candidates, 7 of which were 
confirmed as members of Taurus\footnote{\citet{luh04tau} incorrectly stated 
that 17 candidates were selected with the methods described in this work, when 
in fact the number was 15; the two other candidates, 2MASS~04185791+2830520 
and 04305971+1804237, were selected from photometry in \citet{luh00tau}.}.
Spectroscopy for the other 68 candidates is described in the next section.
In \S~\ref{sec:spec}, I classify 22 candidates as Taurus members and 
46 candidates as field stars.

\begin{figure}
\plotone{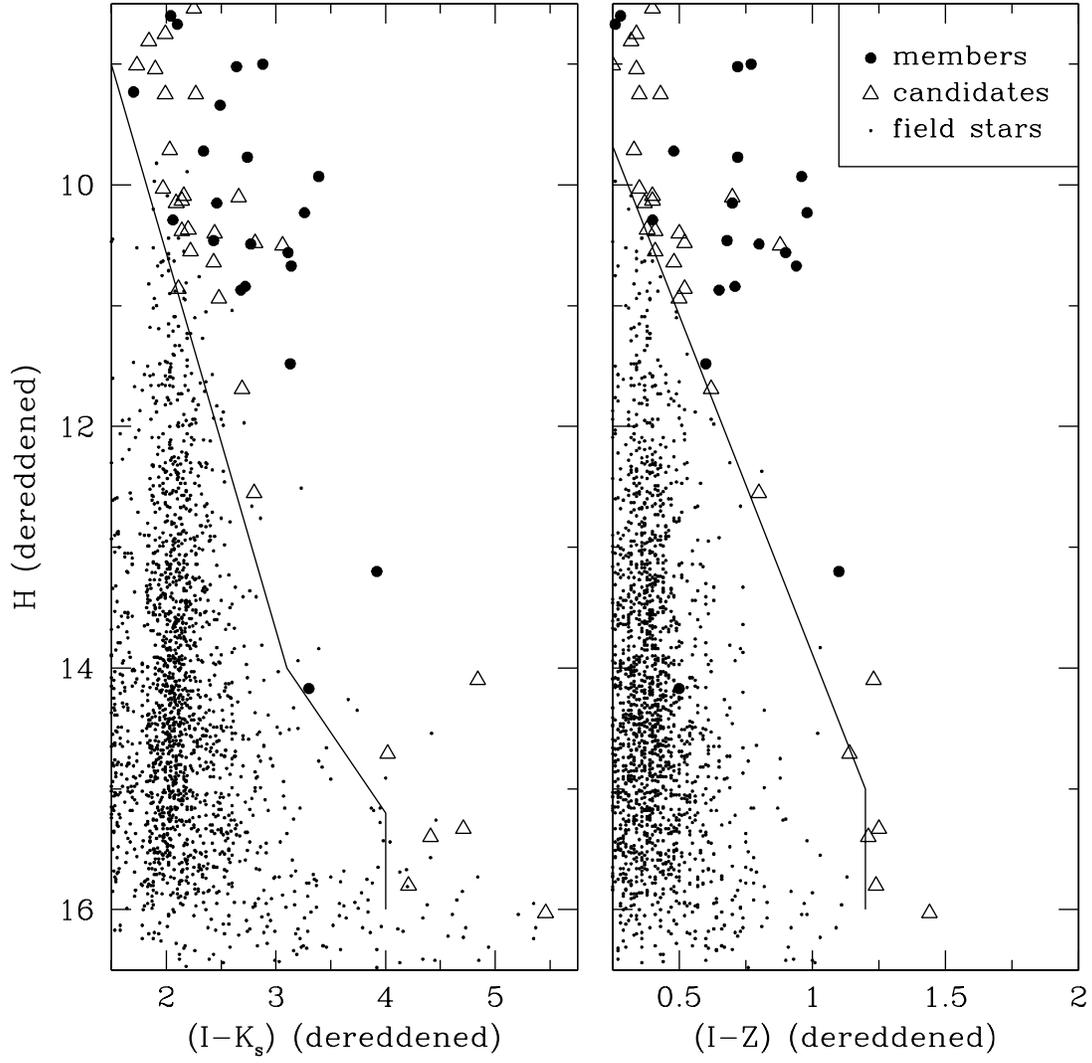}
\caption{
Extinction-corrected color-magnitude diagrams for stars with $A_V\leq8$ in
the Taurus survey from \citet{luh00tau}. 
Stars above both of the solid boundaries are candidate members of Taurus
({\it triangles}) while stars below either of the boundaries are 
likely to be field stars ({\it small points}).
The seven faintest candidates have been observed spectroscopically in this work.
}
\label{fig:4sh1}
\end{figure}

I also examined the photometry from the surveys by \citet{luh00tau} and 
\citet{bri02} for candidate low-mass members of Taurus that have not
been observed spectroscopically. In Figure~\ref{fig:4sh1},
the data from \citet{luh00tau} are shown in
extinction-corrected diagrams of $I-K_s$ versus $H$ and $I-z\arcmin$ versus $H$,
which were constructed in the manner described by \citet{luh04cha}.
I include with the data the boundary for separating potential members of 
Taurus from probable field stars that was developed by \citet{luh04tau}.
Several objects appear above the boundaries of both diagrams, and
thus are candidate members.  Few undiscovered members are expected to reside 
among the brighter candidates ($H<12$) since the survey of \citet{bri98} 
already searched those magnitude levels for the fields in question. 
However, seven faint sources ($H>12$) are above the boundaries of both 
diagrams, and thus are candidate substellar members.
I obtained spectra of these seven candidates, as well as two similar 
candidates from the data of \citet{bri02}. These nine candidates 
are classified as field stars in \S~\ref{sec:spec}.

In addition to the candidate members of Taurus identified in this section,
I selected for spectroscopy three known members that lack accurate spectral 
classifications (IRAS~04370+2559, IRAS~04166+2706, CIDA~7),  
three members that were classified as low-mass class~I objects
by \citet{whi04} (IRAS~04158+2805, IRAS~04248+2612, IRAS~04489+3042), 
and nine new members from G06 (CFHT~5, 7, 11, 15-18, 20, 21).

\section{Spectroscopy of Candidates}
\label{sec:spec}

\subsection{Observations}

I performed spectroscopy on the 77 candidate Taurus members 
and the 15 known members that were selected in the previous section.
Optical and near-IR spectra were obtained for 65 and 18 of the candidates and 
14 and 4 of the known members, respectively. 
Table~\ref{tab:log} summarizes the observing runs and instrument configurations
for these data. In Tables~\ref{tab:field} and \ref{tab:mem}, I indicate the
night on which each object was observed.
The procedures for the collection and reduction of the optical spectra
with the MMT and Hobby-Eberly Telescope (HET) were similar to those described 
by \citet{luh04tau}.  The IR spectra obtained with SpeX \citep{ray03} at the 
Infrared Telescope Facility (IRTF) were reduced with the Spextool package 
\citep{cus04} and corrected for telluric absorption with the method from 
\citet{vac03}.

\subsection{Spectral Classification}
\label{sec:class}

To measure spectral types and assess membership in Taurus for the objects 
in my spectroscopic sample, I applied the optical and IR classification methods 
from my previous studies of Taurus and other star-forming regions 
\citep{luh99,luh04cha,luh05flam}.  
The spectral types are based predominantly on the absorption bands of VO and 
TiO ($\lambda<1.3$~\micron) and H$_2$O ($\lambda>1$~\micron).
When classifying objects that appear to be young,
averages of spectra of dwarfs and giants are used as the spectroscopic 
standards at optical wavelengths \citep{luh99}. 
For the IR spectra, optically-classified young objects are used as the 
standards, which ensures that the IR and optical types are on the same 
classification system. If dwarfs were instead used as the standards
for classifying the IR spectra of Taurus members, the resulting spectral
types would be systematically later than those derived from the optical
spectra because the H$_2$O absorption bands are stronger 
in pre-main-sequence sources than in field dwarfs at a given optical
spectral type \citep{lr99,luc01,mc04}.
To determine whether an object is a member of Taurus or a field star, 
I employ the diagnostics described in my previous work 
\citep[e.g.,][]{luh03b,luh05flam,luh04cha}, such as emission lines,
IR excess emission, gravity-sensitive spectral features, 
and reddening. Only a few of the spectra have sufficiently high resolution
and signal-to-noise for a reliable measurement of Li absorption, which is
another indicator of youth.

\begin{figure}
\epsscale{.9}\plotone{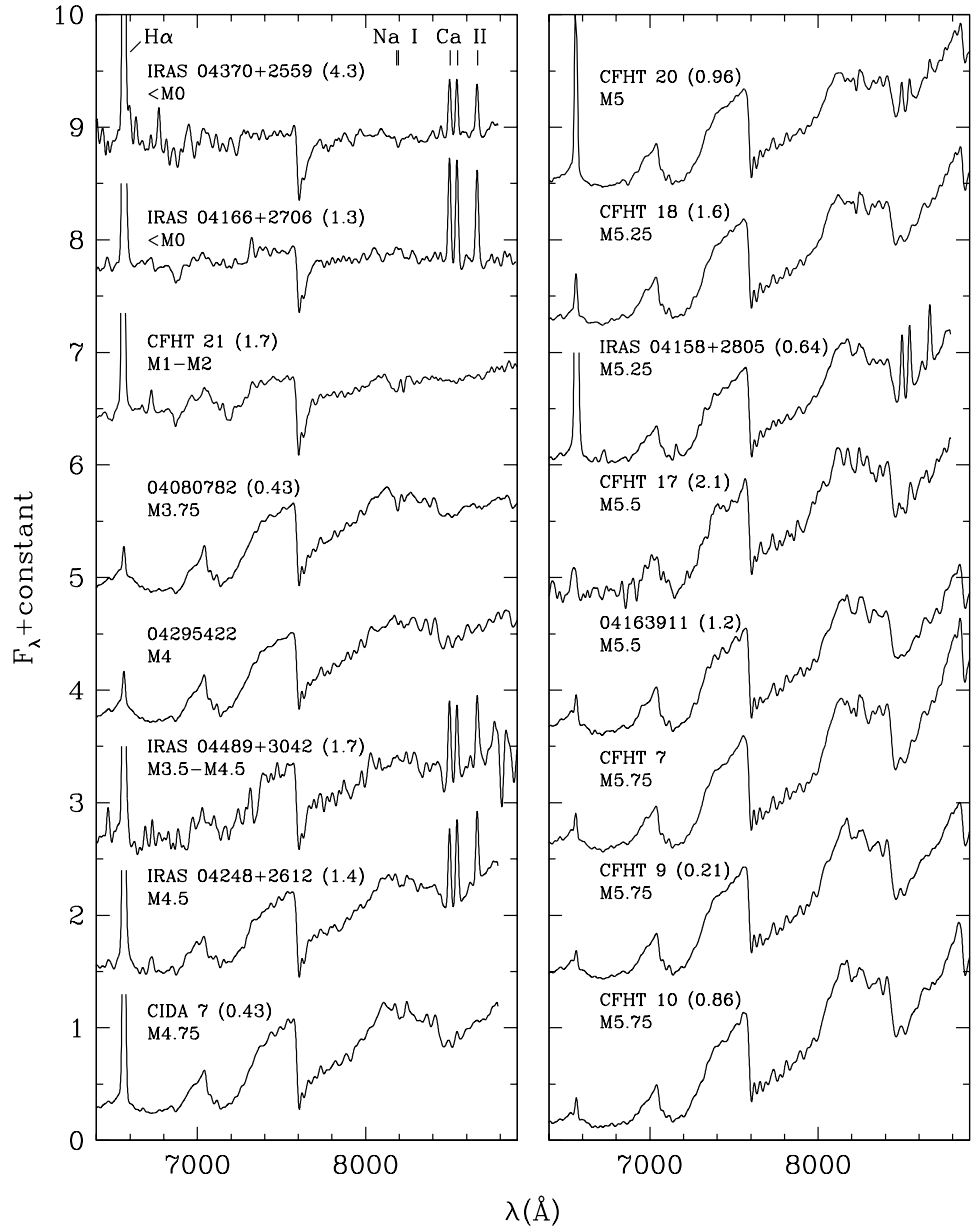}\epsscale{1}
\caption{
Optical spectra of previously known members of the Taurus star-forming region
(IRAS and CFHT) and new members identified in this work
(eight digit identifications).
The spectra have been corrected for extinction, which
is quantified in parentheses by the magnitude difference of the reddening
between 0.6 and 0.9~\micron\ ($E(0.6-0.9)$).
The data are displayed at a resolution of 18~\AA\ and are normalized at
7500~\AA.
}
\label{fig:op1}
\end{figure}

\begin{figure}
\plotone{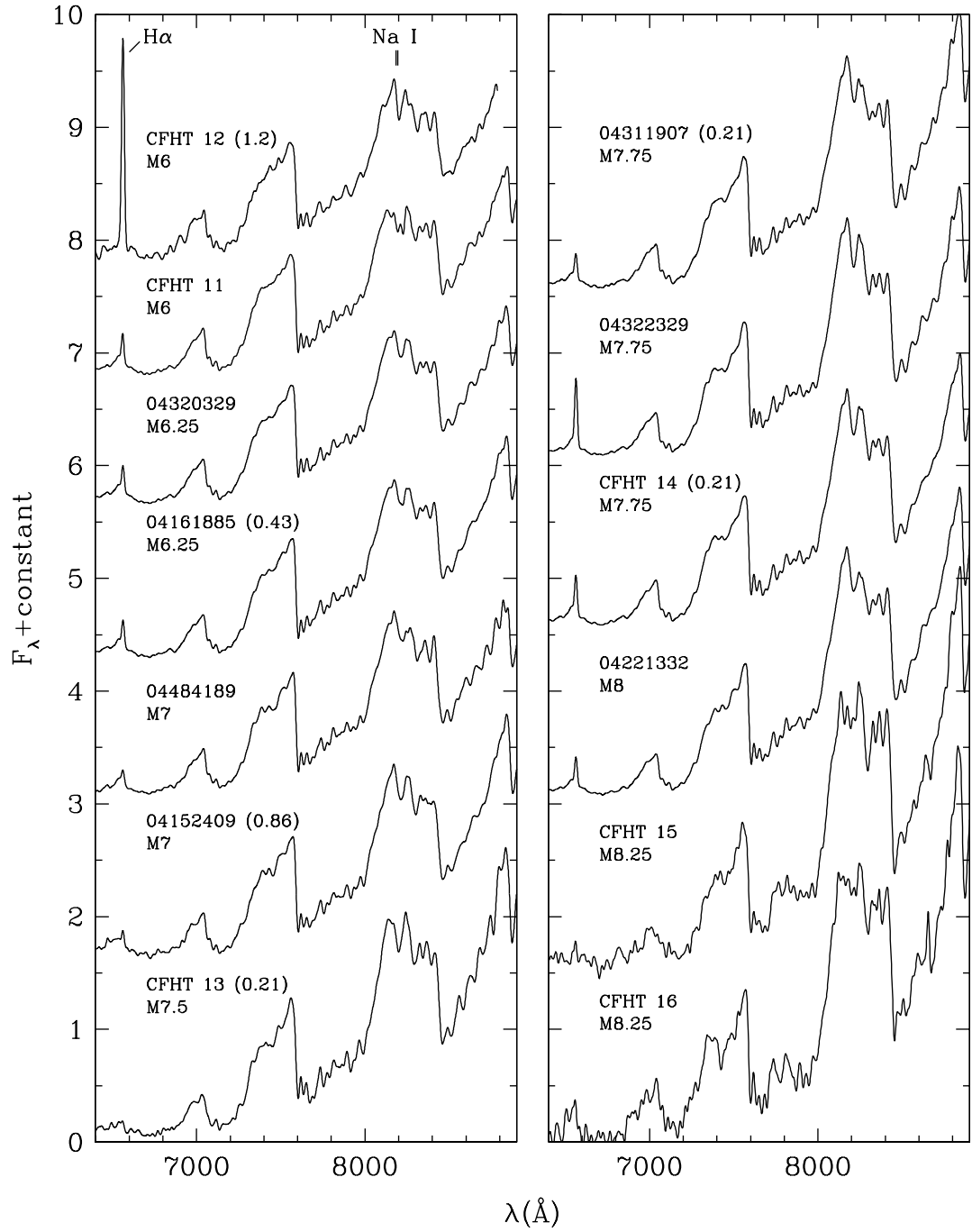}
\caption{
More optical spectra of previously known and new members of Taurus
(see Fig.~\ref{fig:op1}).
}
\label{fig:op2}
\end{figure}

\begin{figure}
\plotone{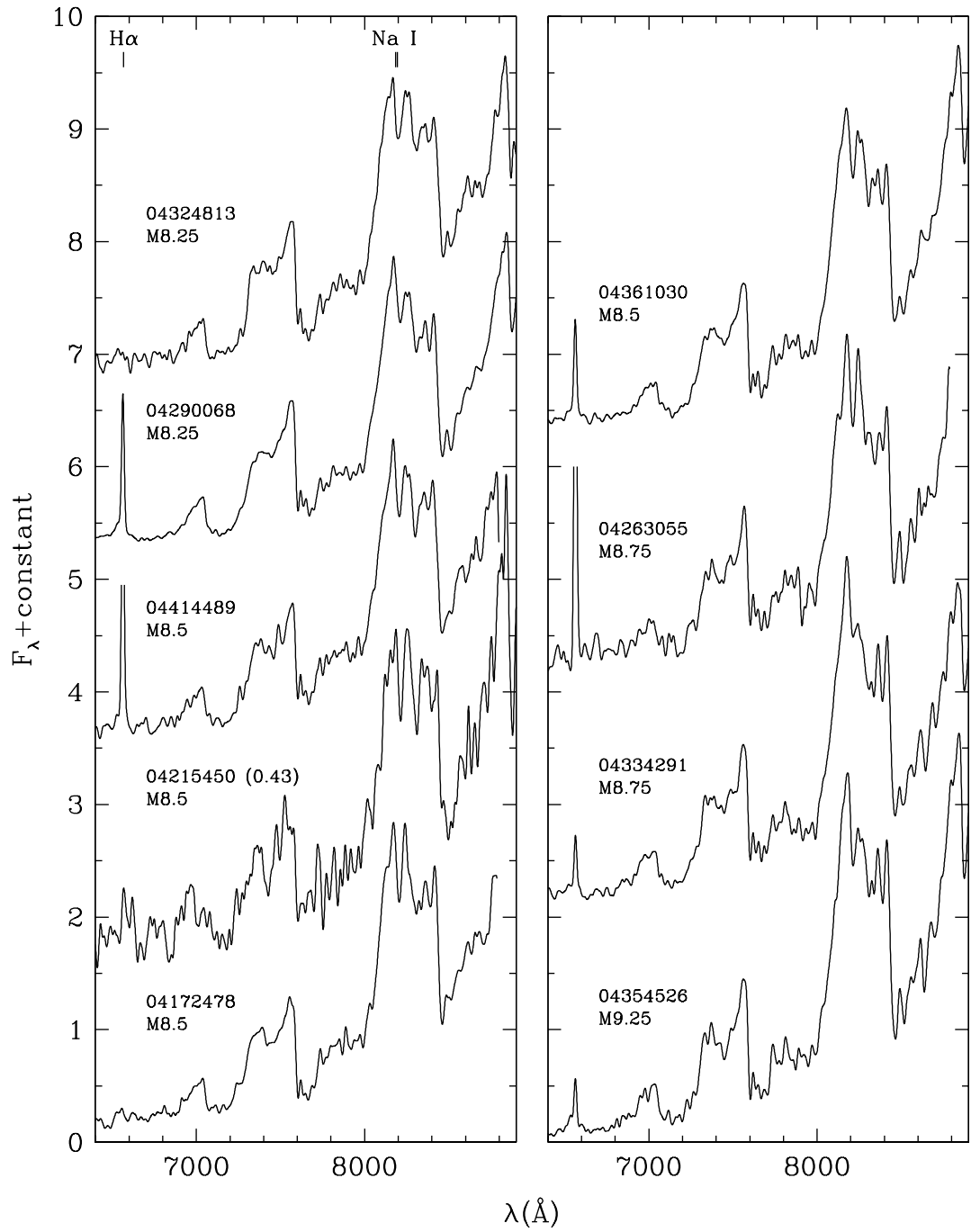}
\caption{
More optical spectra of known and new members of Taurus 
(see Fig.~\ref{fig:op1}) and two objects 
(04324813, 04172478) that may be young members of the 
field rather than Taurus members (see Figs.~\ref{fig:ir1} and \ref{fig:ir2}).
}
\label{fig:op3}
\end{figure}

\begin{figure}
\plotone{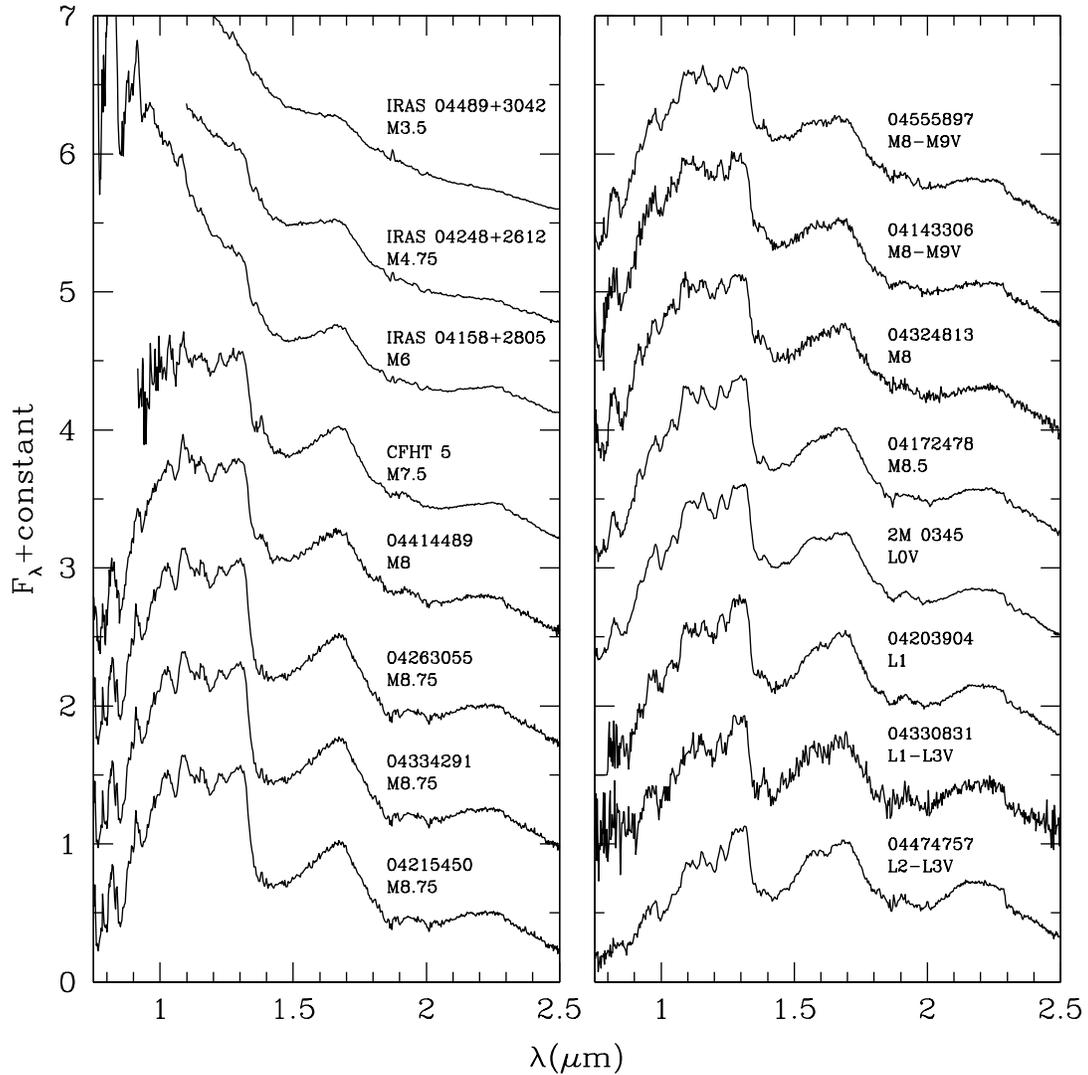}
\caption{
Near-IR spectra of known Taurus members (IRAS and CFHT) 
and candidate members identified in this work
(eight digit identifications).
Candidates that exhibit evidence of youth (and hence membership in Taurus)
in the form of triangular $H$-band continua are shown 
with the known members ({\it left}). 
For the other candidates ({\it right}), the shapes of the $H$-band continua 
are dwarf-like (04555897, 04143306, 04330831, 04474757) or intermediate 
between dwarfs and Taurus members (04324813, 04172478, 04203904). 
A spectrum of the field L dwarf 2M0345+25 is included among the field stars
for comparison.
The spectra for the young objects on the left have been dereddened to 
the same slope as measured by the ratios of fluxes at 1.32 and 1.68~\micron.
These data have a resolution of $R=100$ and are normalized at 1.68~\micron.
}
\label{fig:ir1}
\end{figure}

Based on the optical and IR spectra, I classify 55 candidates as 
field stars and 22 candidates as members of Taurus. 
Five of the 22 new members (CFHT~9, 10, 12-14) were independently 
discovered in the recent survey by G06.
The identifications and photometry from 2MASS and the available 
spectral types for these field stars and new members are provided in 
Tables~\ref{tab:field} and \ref{tab:mem}, respectively. 
Table~\ref{tab:mem} also includes the 15 previously 
known members that were observed spectroscopically, as well as the evidence 
of membership for each object. 
The spectra of the known and new members are presented in order of 
spectral type in Figs.~\ref{fig:op1}-\ref{fig:ir1}. 
To facilitate the comparison of these spectra, they have been 
corrected for reddening \citep{luh04cha,luh05flam}.
The positions of the new members are indicated in a map of Taurus 
in Figure~\ref{fig:map1}.
Interestingly, one of the new brown dwarfs, 
2MASS~04221332+1934392, is only $4\arcmin$ from T~Tau, which 
has only one other known young star within $2\arcdeg$.

The studies cited above have provided many illustrations of the 
variations of the gravity-sensitive spectral features between 
dwarfs, giants, and pre-main-sequence objects that appear in the kind of 
optical and IR spectra obtained in this work.  
I present an additional illustration for the IR spectra
by including in Figure~\ref{fig:ir1} the spectra of the candidates 
classified as late-type field dwarfs.
In the field dwarfs, the $H$-band continua exhibit broad plateaus,
whereas the Taurus members are characterized by sharply peaked, triangular 
continua. This behavior has been observed in young late-type objects
in other star-forming regions \citep{luc01,luh04ots} and has been
attributed to the dependence of H$_2$ collision induced absorption on 
surface gravity \citep{kir06}. 
A similar effect is found in the $K$-band, although it is more subtle than
at $H$. Another feature that varies noticeably in these IR spectra is 
the FeH absorption at 0.99~\micron\ \citep{mc04,kir06}, which is strong in the 
dwarfs and very weak or undetected in the young objects.

\subsection{Comments on Individual Sources}

\begin{figure}
\plotone{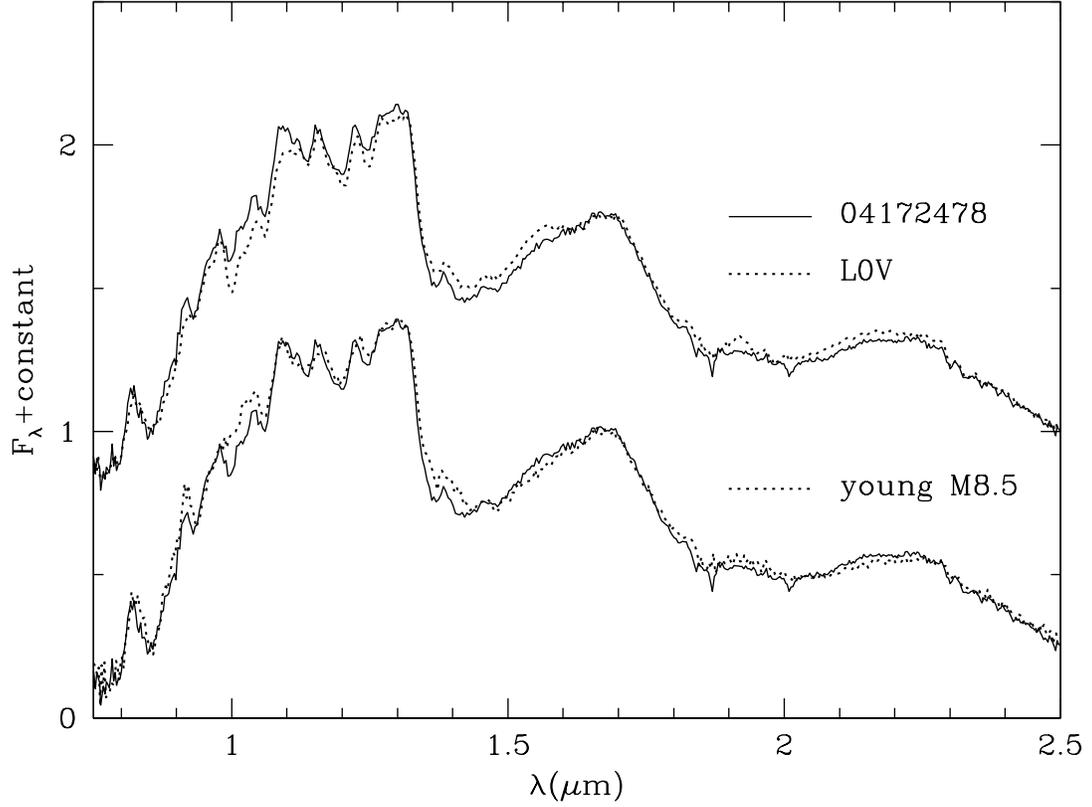}
\caption{
Near-IR spectrum of 2MASS~04172478+1634364 ({\it solid line}) 
compared to data for a field dwarf (2M~0345, L0V) and a late-type member of
Taurus (KPNO~6, M8.5) ({\it dotted lines}).
For  2MASS 04172478+1634364,
the strength of the FeH absorption at 0.99~\micron\ and the shapes of the
$H$- and $K$-band continua are intermediate between 
those of the field dwarf and the Taurus member, indicating an intermediate 
surface gravity and age. An object not shown here, 2MASS 04324813+2431063,
exhibits similar characteristics. 
These data have a resolution of $R=100$ and are normalized at 1.68~\micron.
}
\label{fig:ir2}
\end{figure}

\begin{figure}
\plotone{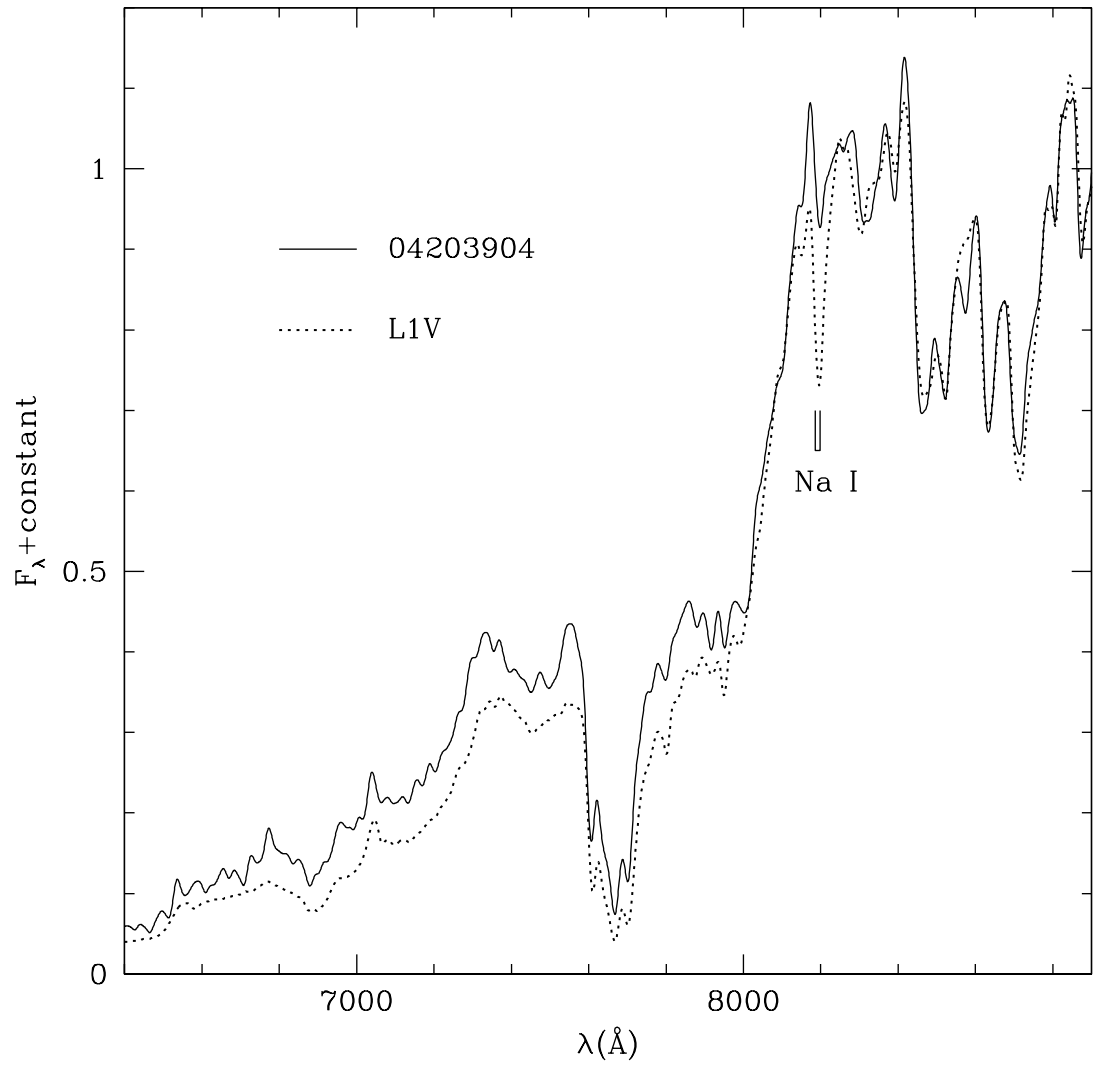}
\caption{
Optical spectrum of 2MASS~04203904+2355502 ({\it solid line}) compared to 
the spectrum of an L1 field dwarf 
\citep[2MASS~1439284+192915, {\it dotted line},][]{kir99}. 
2MASS~04203904+2355502 appears to have a lower surface gravity than the
dwarf based on its weaker Na~I absorption.
The data are displayed at a resolution of 18~\AA\ and are normalized at
8250~\AA.
}
\label{fig:op4}
\end{figure}

In terms of the gravity-sensitive features, most of the candidates are 
well-matched to either known Taurus members or standard field dwarfs.
The exceptions are 2MASS~04172478+1634364, 04324813+2431063,
and 04203904+2355502.
The IR spectrum of one of these objects, 2MASS~04172478+1634364, is compared
to data for a Taurus member and a field dwarf in Figure~\ref{fig:ir2}.
Both the FeH absorption and the shape of the $H$-band continuum are intermediate
between those of the comparison objects, which indicates an intermediate
surface gravity.
The other two objects exhibit the same behavior, except that the
FeH absorption of 2MASS~04203904+2355502 is dwarf-like. 
The gravity-sensitive lines at optical wavelengths are consistent
with surface gravities lower than those exhibited by field dwarfs for these
three sources. For instance, in addition to the spectra of known members, 
Figure~\ref{fig:op3} includes the optical spectra of 
2MASS~04172478+1634364 and 04324813+2431063,
whose Na~I and K~I strengths are indistinguishable from those of the
Taurus members. Similarly, the third object, 
2MASS~04203904+2355502, exhibits much weaker Na~I than a standard field L 
dwarf \citep{kir97}, as shown in Figure~\ref{fig:op4}.
These optical and IR data are consistent with ages older than Taurus 
($\tau>2$~Myr) and younger than typical field dwarfs ($\tau<1$~Gyr). 
These three sources may represent low-mass members of the population of young 
field stars ($\tau\sim100$~Myr) that \citet{bri97} used to explain
the wide-field distribution of X-ray sources in the direction of Taurus. 
In this case, 2MASS~04203904+2355502 could be a young field L dwarf like 
the one recently discovered by \citet{kir06}.
Because of the uncertainty in their ages, it is unclear whether to use
dwarfs, Taurus members, or stars at some other age as standards when 
measuring their spectral types. For 2MASS~04172478+1634364 and 04324813+2431063,
the use of dwarfs as standards produces late-M spectral types from the 
optical spectra and early L types from the IR data. Meanwhile, similar
spectral types are derived from the two wavelength ranges when Taurus
members are used as the standards. Therefore, I adopt the latter types
for the purposes of this work. For 2MASS~04203904+2355502, both the optical
and IR spectra are consistent with the same type of L1 when compared to
dwarf standards, which is adopted here.

\begin{figure}
\plotone{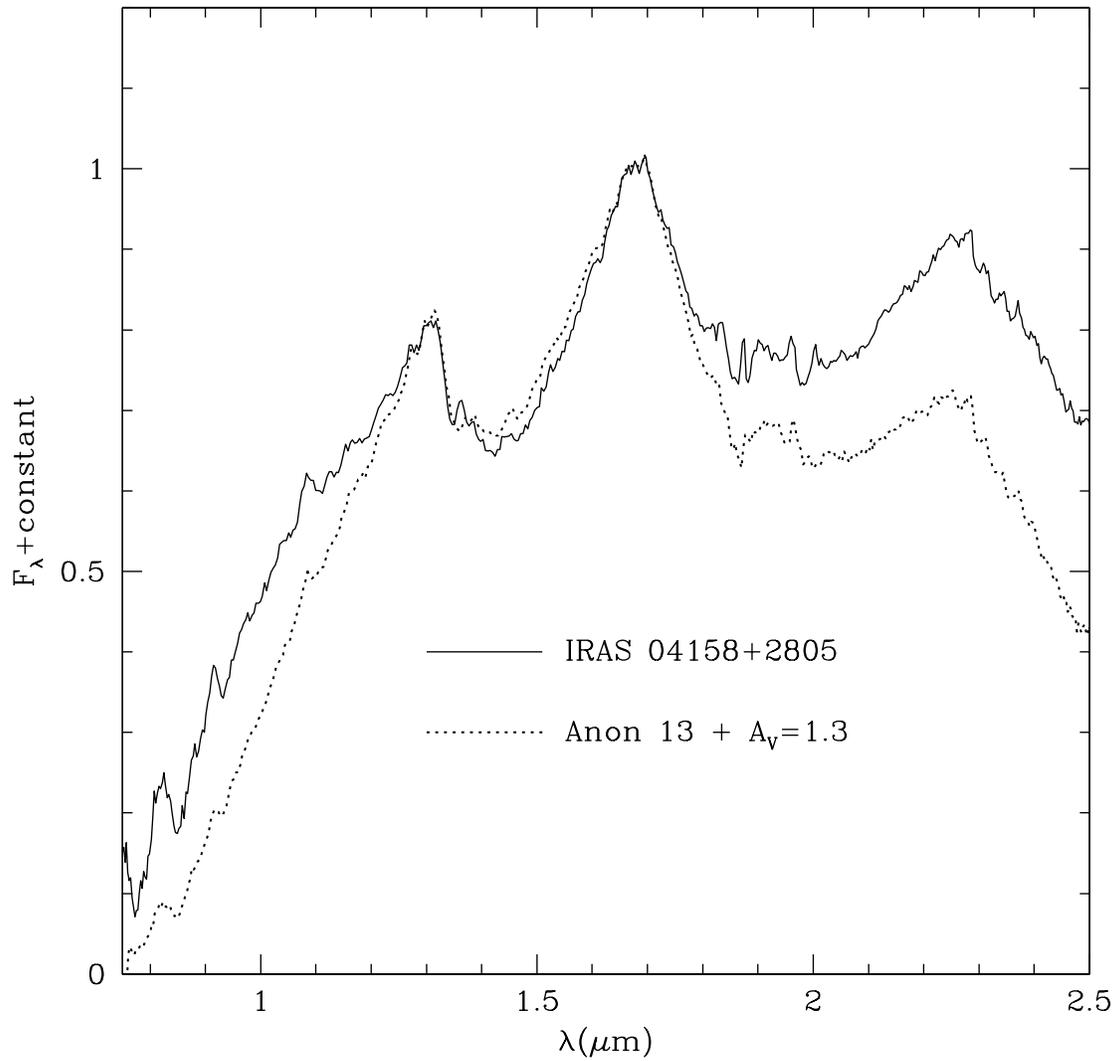}
\caption{
Near-IR spectrum of IRAS~04158+2805 ({\it solid line}) compared to a
spectrum of V410~Anon~13 \citep[][{\it dotted line}]{luh05wfpc}.
The spectral type of IRAS 04158+2805 is similar to that
of V410~Anon~13 based on the strengths of the steam absorption bands.
IRAS~04158+2805 is brighter than V410 Anon 13 at both short and 
long wavelengths, possibly due to excess emission from accretion and 
a circumstellar dust, respectively.
The spectrum of V410~Anon~13 has been reddened to match the data for 
IRAS~04158+2805 at 1.32 and 1.68~\micron.
These data have a resolution of $R=100$ and are normalized at 1.68~\micron.
}
\label{fig:ir3}
\end{figure}

In addition to candidate members of Taurus, several previously known
members were included in my spectroscopic sample. I briefly comment on
the spectra of three of these known members, 
IRAS~04158+2805, IRAS~04248+2612, and IRAS~04489+3042, which were
classified as possible class~I brown dwarfs by \citet{whi04}. 
The optical spectra of these objects have strong emission in H$\alpha$, 
Ca~II, and other permitted and forbidden lines. For all three sources, 
both the optical (0.6-0.9~\micron) and near-IR
(0.75-2.5~\micron) spectra exhibit increasing excess emission with decreasing 
wavelength at $\lambda<1$~\micron\ when compared to the classification
standards. Long wavelength excess emission at $>$2~\micron\ is also apparent
in the data for IRAS~04158+2805.  In Figure~\ref{fig:ir3}, 
the blue and red excesses for this star are revealed by a comparison to 
another Taurus member at the same spectral type, V410~Anon~13.
To avoid veiling by excess emission, the 
spectral types for these stars are primarily based on features at
$\lambda>0.8$~\micron\ and $\lambda=1$-2~\micron\ in the optical and
near-IR spectra, respectively.  For each of the three objects, the optical 
and near-IR spectra produce similar spectral types. 
However, the optical spectra imply much lower extinctions ($\Delta A_V=5$-15)
than the IR spectra, which is probably caused by the blue excess emission. 
The spectral types measured here for IRAS~04158+2805, IRAS~04248+2612, and 
IRAS~04489+3042 are earlier than the ones reported by \citet{whi04},
but the two sets of classifications are consistent within the uncertainties. 
The combination of my spectral types, the
temperature scale of \citet{luh03b}, and the evolutionary models of 
\citet{bar98} and \citet{cha00} suggest that these three class~I sources
are low-mass stars rather than brown dwarfs.

\section{Survey Completeness}
\label{sec:complete}

All of the fields in Taurus that have been previously searched for new 
low-mass members are encompassed by the 225~deg$^2$ area
considered in this work.
Thus, the completeness of the various surveys can be investigated
in a straightforward manner by comparing the lists of late-type objects
discovered in each survey. In this section, I focus on the completeness
of my survey and the recent 28~deg$^2$ survey by G06.

I first examine the completeness of my USNO/2MASS survey for the population 
that was targeted, namely Taurus members with spectral types of $>$M6.
In this discussion, I include the initial sample of members found in this 
survey that was reported in \citet{luh04tau}. 
As mentioned in \S~\ref{sec:cand}, two of the 14 previously known late-type 
members of Taurus from \citet{luh03tau} and prior studies, KPNO~9 (8.5) and 
KPNO~12 (M9), have 2MASS photometric uncertainties that are greater than 
the thresholds I adopted for selecting candidates.
Three of the six new members at $>$M6 that were found in the CCD survey by 
\citet{luh04tau} did not satisfy the criteria for the USNO/2MASS survey 
because ITG2 (M7.25) is slightly too bright at $H$, 
2MASS~04552333+3027366 (M6.25) is too blue in $I2-K_s$, and 
2MASS~04574903+3015195 (M9.25) has photometric uncertainties that are too large.
Among the 6 new members from G06 that I classify as $>$M6, 
three objects were not found in my survey because two of them
are slightly too faint (CFHT~15, 16) and one is too red in $J-H$ (CFHT~5), 
and thus is outside of the extinction range considered here.
These results suggest that my survey has a completeness of $\sim75$\% for
Taurus members at $>$M6, $\leq$M9, and $A_V\leq4$. 
The completeness would have been $>90$\% if I had selected a slightly
brighter limit at $H$ and slightly larger thresholds for the 2MASS
photometric uncertainties (e.g., 0.06~mag instead of 0.05~mag).

Before discussing the completeness of the wide-field survey by \citet{mar01} 
and G06, I summarize its results and place it in the 
context of other surveys of Taurus.
\citet{mar01} obtained spectra of candidate low-mass members
appearing in an initial set of images covering 3.6~deg$^2$,
resulting in the discovery of four new members (CFHT~1-4). 
G06 then presented the results of their full
survey of several fields encompassing a total area of 28~deg$^2$. 
After applying their selection criteria for identifying candidate low-mass
members, they recovered 17 previously known members of Taurus.
G06 performed spectroscopy on eight of these known members
and treated two of them as new discoveries (CFHT~6 and 8), which
had been found earlier by \citet{luh04tau}.
Through spectroscopy of a large sample of candidates, 
G06 discovered an additional 15 new members. Five of those new objects
have been independently found in this work (CFHT~9, 10, 12-14). 
Among the other 10 new members, six objects were not found in my survey
because they have spectral types of $\leq$M6 according to my
classifications, and thus are outside of the range of spectral types for which 
my search criteria were designed. The one new member from G06 that was 
not included in my spectroscopic sample, CFHT~19, also has a type earlier than
M6 according to G06. The other three new CFHT sources that 
were not found in my survey were discussed earlier in this section.
Using their spectroscopy, G06 classified 12 of their 17 
new members as $>$M6, and thus likely to be substellar. 
In comparison, the spectral types measured in this work for those sources
are earlier than the classifications from G06
by an average of 0.5 subclass. As a result, I find that only 6/17 of the
CFHT sources from G06 are later than M6.

I now evaluate the completeness of the survey by G06.
They identified 47 objects with photometry indicative 
of Taurus members with spectral types of $\geq$M4.
These sources included the four members presented by \citet{mar01} and 
17 previously known members. G06 obtained spectra of 20 of the 26 
remaining candidates, which included all of the candidates at $>$M6.
As a result, they reported a completeness of 41/47, or 87\%, for 
Taurus members at $\geq$M4 and a completeness of 100\% for 
candidates at $>$M6, both for $M>0.03$~$M_\odot$, $A_V\leq4$, and ages
of $\tau\leq10$~Myr.
However, these percentages do not fully characterize completeness 
because they did not account for the completeness of the candidate list 
itself, or the percentage of members that were identified as candidates, 
which G06 did not estimate.
As a simple test of their completeness, I have investigated whether  
any known members of Taurus with spectral types of $\geq$M4 are
within their survey fields but not recovered by their selection criteria. 
From published membership lists for Taurus 
\citep{hb88,ss94,kh95,bri98,bri99,bri02,mar00,mar01,lr98,luh03tau,luh04tau},
I have identified 38 such sources: 
MHO~4-9, V927~Tau, RXJ04467+2459, CIDA~1, 2 and 7, 
J1-507, J1-665, J1-4423, IRAS~04248+2612 and 04158+2805, 
LkCa~1, V410~X-ray~1, 3, 4, 5a, and 6, FQ~Tau, 2MASS~04161210+2756385,
04213459+2701388, 04403979+2519061, 04141188+2811535, and 04414825+2534304,
KPNO~8, 13 and 14, UX~Tau~C, Haro~6-32, FN~Tau, GG~Tau~Ba, FW~Tau, GM~Tau, and 
ITG~2. Seven sources of this kind are also present among the 
new members discovered in this work: 
2MASS~04263055+2443558, 04334291+2526470, 04320329+2528078, 04322329+2403013,
04311907+2335047, 04215450+2652315, 04295422+1754041.
Among these 45 members, 34 and 11 have spectral types of M4-M6 and $>$M6,
respectively. In comparison, the total sample of members recovered by 
\citet{mar01} and G06 (both previously known and new ones) consisted 
of 17 and 19 objects in those spectral type ranges. Thus, the completeness of 
G06 was at best 17/51 (33\%) and 19/30 (63\%) at M4-M6 and $>$M6.
The true completeness at M4-M6 is probably lower because neither 
G06 nor any other survey has demonstrated that the known census of 
members in this range of spectral types is complete for the area considered
by G06. On the other hand, their completeness at $>$M6 may not
be much lower than the upper limit computed here given that my survey 
was designed to find members of this kind, and has a reasonably high level 
of completeness for M6-M9 and $A_V\leq4$, as shown earlier in this section. 

\begin{figure}
\plotone{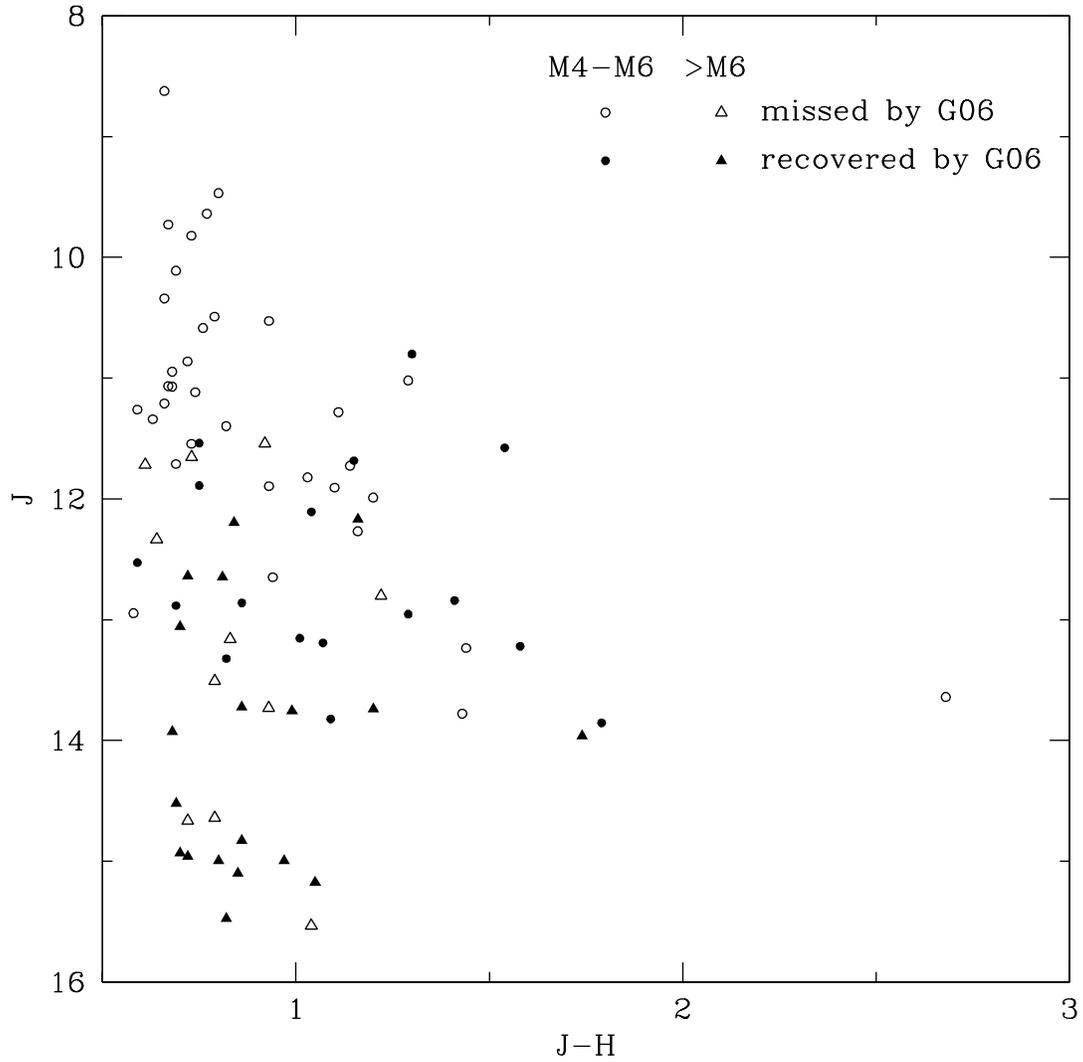}
\caption{
$J$ versus $J-H$ for known Taurus members with spectral types of M4-M6
and $>$M6 that were recovered ({\it solid circles and triangles}) 
and that were missed ({\it open circles and triangles}) by the 
survey of \citet{gui06}. The latter are systematically brighter than the 
former, suggesting that most of them were missed because of saturation.
}
\label{fig:jh}
\end{figure}

The reason for the high level of incompleteness in the survey 
by G06 is explored in Figure~\ref{fig:jh}, which compares
$J$ versus $J-H$ for the 36 recovered members and the 45 missed members.
All but one of the missed members span a range of colors that is similar
to, or even smaller on average than, the recovered members, 
indicating that the former objects were not missed because of higher 
extinction. Instead, the missed members are systematically brighter than
the recovered members, and thus may have been saturated in the images of 
G06 and consequently overlooked.

\section{Spatial Distribution of Brown Dwarfs}
\label{sec:spatial}

My 225~deg$^2$ survey of Taurus was designed to identify members 
in the same ranges of spectral types (M6-M9) and extinctions
($A_V\leq4$) as exhibited by the previously known late-type members that were 
found in surveys of smaller fields.
As shown in the previous section, the completeness of this survey is not
100\% for those ranges of types and extinctions, but it appears to be 
high enough that the combination of the new members that I have found and
the previously known members should comprise an accurate representation 
of the substellar population of Taurus that is unbiased with position.
In this section, I examine the spatial distribution of that population.

A map of the positions of all known members of Taurus is shown in 
Figure~\ref{fig:map2}, where the objects with spectral types of
$\leq$M6 ($M\gtrsim0.1$~$M_\odot$) and $>$M6 ($M\lesssim0.08$~$M_\odot$)
are plotted with different symbols.
A large, extended population of brown dwarfs between the 
stellar aggregates does not exist.
Instead, the spatial distribution of brown dwarfs in Taurus closely
resembles that of the stellar members, which is consistent with previous
surveys of smaller areas \citep[][G06]{bri02,luh04tau}.

To search for subtle differences in the distributions of stars and
brown dwarfs in Taurus in their surveyed fields, G06 computed 
the number ratio of brown dwarfs to stars for areas of varying radii
from the stellar aggregates, which is quantified as 
${\mathcal R}_{1}=N(0.02\leq M/M_\odot\leq0.08)/N(0.08<M/M_\odot\leq10)$
\citep{bri02}. They found that this ratio increased with radii. For instance,
G06 measured ${\mathcal R}_{1}=0.23$ for their full survey area, 
whereas \citet{luh04tau} reported a value of 0.18 for smaller fields
surrounding some of the aggregates. G06 presented these results as 
evidence for differences in the spatial distributions of stars and 
brown dwarfs in Taurus. However, as discussed in \S~\ref{sec:complete},
the spectral types reported by G06 for the CFHT sources
are systematically later than the ones that I measured for those objects.
If my spectral types are adopted instead, which are
based on the classification system used for most of the
previously known late-type members \citep{bri02,luh03tau,luh04tau},
then six CFHT objects become stars rather than brown dwarfs 
($\leq$M6, CFHT~7-12)
and the ratio computed by G06 becomes 0.17, which
is consistent with the value for the smaller fields. In other words, 
the apparent increase in ${\mathcal R}_{1}$ with radius reported
by G06 was simply a reflection of differences in the 
spectral classification systems used for their (outer) CFHT sources 
and the (inner) previously known members. 
However, even if the measurement of ${\mathcal R}_{1}$ by G06 is
revised to use my spectral types for the CFHT sources, it still may not 
accurately represent the Taurus population because,
as shown in the previous section, their survey 
was significantly incomplete for members at $\geq$M4, which applies to 
both the numerator and denominator in ${\mathcal R}_{1}$ and thus 
precludes a meaningful ratio.

Although the combination of previous surveys and the one in this work
provide a census of brown dwarfs that has fairly high completeness for
M6-M9 and $A_V\leq4$ across all of Taurus, I do not attempt to measure 
${\mathcal R}_{1}$ for the entire star-forming region because no study
has quantified the completeness of the known census of low-mass stars (M2-M6)
for this area. This issue was raised by \citet{luh04tau}, 
who pointed out that the faint limits of the wide-field X-ray and H$\alpha$ 
surveys ($\sim$M4) do not clearly overlap with the bright limits of 
deep CCD surveys, which range from M4 to M6 
\citep[][G06]{bri98,bri02,luh00tau,luh04tau}.
For instance, one of the objects found in my survey, 2MASS~04161885+2752155 
(M6.25), is within one of the fields observed by \citet{bri02}, but was
not found in that study because it was saturated. 
Paradoxically, the membership of Taurus now is probably more complete for
brown dwarfs than for low-mass stars because the latter have been too faint for 
X-ray surveys, frequently saturated in CCD surveys, and not easily 
distinguished from background field stars in USNO and 2MASS data.

\begin{figure}
\plotone{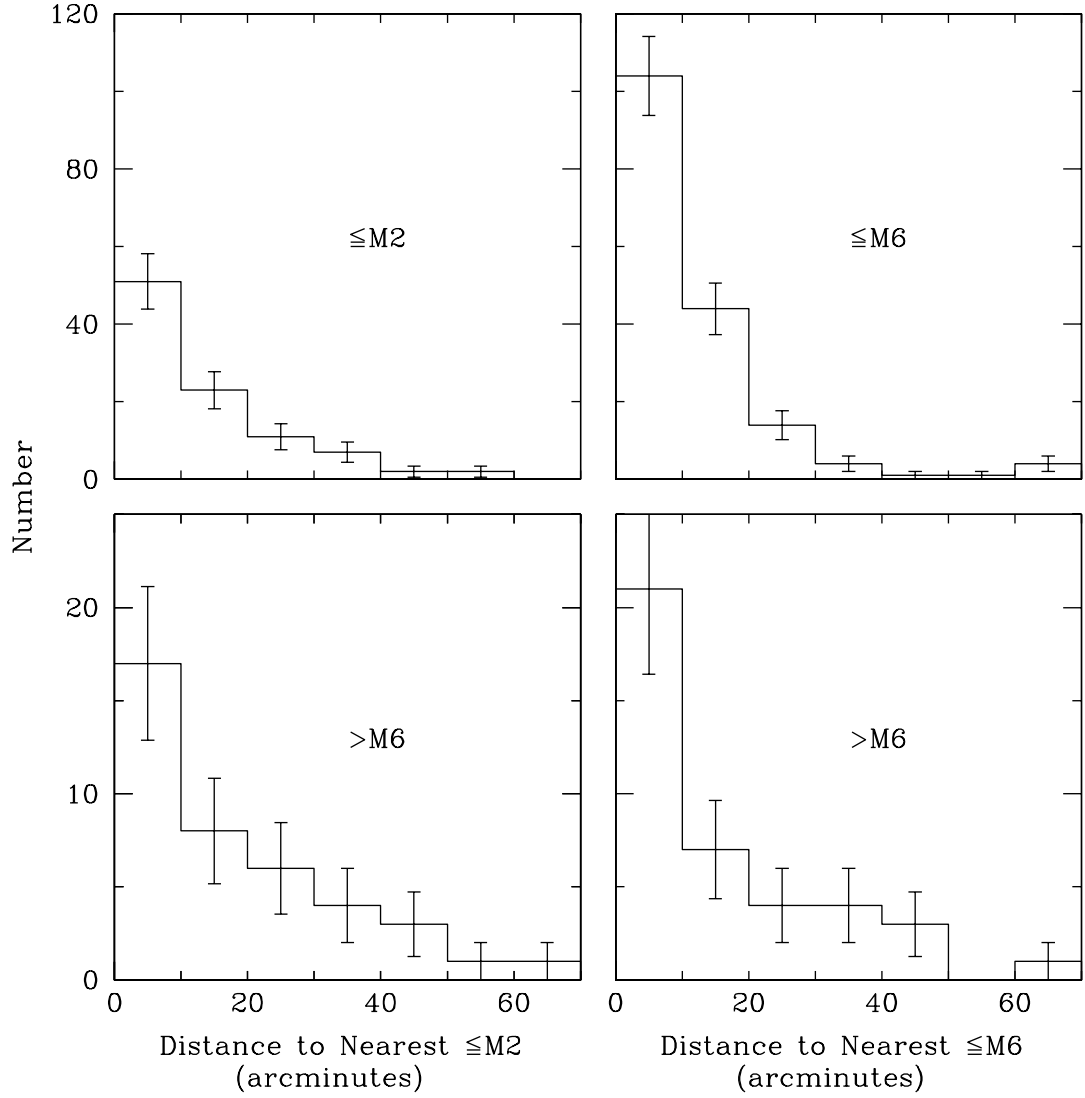}
\caption{
Distributions of projected angular distances among members of Taurus. 
{\it Left}:
For substellar members of Taurus ($>$M6, {\it bottom}), 
the distribution of distances to the nearest star at $\leq$M2 is similar 
to the distribution of nearest distances among those stars ({\it top}). 
{\it Right}:
The same result is found when members at M2-M6 are included in the stellar
sample. 
}
\label{fig:histo}
\end{figure}

As demonstrated above, a comparison of a ratio like 
${\mathcal R}_{1}$ with position (or with any other parameter) is
sensitive to incompleteness and differences in spectral classification 
systems. To avoid the latter effect, one can examine the 
spatial distributions of stars and brown dwarfs in terms of the angular 
distances to the nearest stellar neighbors instead of ${\mathcal R}_{1}$.
Changing classification systems is equivalent to changing the boundary
between the two samples in ${\mathcal R}_{1}$, which in turn will change 
quickly because the numerator and denominator are anti-correlated.  
In contrast, a sensitivity of this kind is not present in the distribution 
of neighbor distances because changing the boundary simply removes a few 
objects from one average and adds them to the other average. 
To mitigate the effect of incompleteness on the distribution of 
nearest neighbor distances, I exclude the range of spectral types that
has an uncertain level of completeness, namely M2-M6. 
I have computed the distribution of nearest neighbor distances
for the entire Taurus star-forming region by using the census of Taurus that 
combines the previously known members with the new ones found in this work. 
I consider only members with measured spectral types, which has the effect
of excluding most of the Class~0 and I sources. 
To measure the spatial distributions of unrelated members of Taurus, 
the clustering of members of multiple systems should be avoided in this
experiment. Therefore, neighboring members with separations less than 
$1\arcmin$ are treated as one object \citep{lar95}. The resulting samples 
at $\leq$M2 and $>$M6 contain 101 and 40 sources, respectively.
As shown in Figure~\ref{fig:histo}, these two samples exhibit similar
distributions of distances to the nearest stellar neighbor at $\leq$M2.
A two-sided Kolmogorov-Smirnov test of the two distributions indicates
a probability of $\gtrsim33$\% that they are drawn from the same parent 
distribution, and thus I find no statistically significant difference 
between the spatial distributions of stars and brown dwarfs in Taurus.
I arrive at the same result if a different boundary for the hydrogen 
burning mass limit is adopted, such as M5 or M7, or if I ignore the 
potential incompleteness at M2-M6 and include these members in the
stellar sample (Figure~\ref{fig:histo}).
For their survey area, G06 also found no significant difference
between the distributions of nearest neighbor distances for members
at $\leq$M6 and $>$M6.

\section{Conclusions}

To measure the spatial distribution of substellar members of Taurus, I have 
performed a search for brown dwarfs across the entire star-forming region. 
The results of this work can be summarized as follows:

\begin{enumerate}

\item
Through analysis of photometry from the 2MASS and USNO-B1.0 catalogs
for an area of 225~deg$^2$ encompassing Taurus, I have identified
83 potential substellar members of the region. Spectroscopy was performed
on 15 of these candidates by \citet{luh04tau}, who classified seven of them
as new members. Among the 68 remaining candidates observed in this work, 
I have identified 22 new members, five of which were independently discovered 
by G06.  Sixteen of these 22 objects have spectral types later than M6 and thus
are likely to be brown dwarfs according to the theoretical evolutionary 
models of \citet{bar98} and \citet{cha00} and the temperature scale of 
\citet{luh03b}. 

\item
Among the candidates classified as non-members, three objects have
late spectral types (M8-L1) and spectral features that are suggestive of
surface gravities between those of Taurus members and typical field dwarfs. 
I speculate that these objects could be young members of the field 
($\tau\sim100$~Myr). 

\item
Using the low-mass Taurus members found in previous surveys, I estimate
that my survey has a completeness of $\sim75$\% for members with 
spectral types of $>$M6 to M9 and extinctions of $A_V\leq4$. 
The same type of analysis demonstrates that the completeness of
the recent survey by G06 is $<63$\% for this range of types.

\item
G06 concluded that the abundance of brown dwarfs relative to stars 
varies with distance from the stellar aggregates in Taurus. 
I find that this apparent variation is a reflection of differences in the 
spectral classification systems used for low-mass members inside and outside
of the aggregates. 
No variation is evident when spectral types for all objects are adopted from
the same system. Furthermore, a reliable
measurement of the relative numbers of stars and brown dwarfs as a function
of position in Taurus is currently not possible because the available 
census of Taurus members has an unknown level of completeness for low-mass 
stars (M2-M6, 0.1-0.6~$M_{\odot}$) for most of the region. 

\item
After updating the census of Taurus members with the new objects 
discovered in this work, I find that the spatial distribution of 
brown dwarfs closely follows that of the stars. An extended population of 
brown dwarfs outside of the stellar aggregates is not present.
These results are consistent with a common formation mechanism for stars
and brown dwarfs \citep[e.g.,][]{pn04} and with some models for embryo ejection 
\citep{bat03}, but not others \citep{kb03}.

\end{enumerate}

\acknowledgements

This work was supported by grant NAG5-11627 from the NASA Long-Term Space 
Astrophysics program and was based on observations performed at the
MMT Observatory, the IRTF, and the HET.
The IRTF is operated by the University of Hawaii under Cooperative 
Agreement no. NCC 5-538 with the National Aeronautics and Space 
Administration, Office of Space Science, Planetary Astronomy Program.
The HET is a joint project of the University of Texas at Austin,
the Pennsylvania State University,  Stanford University,
Ludwig-Maximillians-Universit\"at M\"unchen, and Georg-August-Universit\"at
G\"ottingen.  The HET is named in honor of its principal benefactors,
William P. Hobby and Robert E. Eberly.  The Marcario Low-Resolution
Spectrograph at HET is named for Mike Marcario of High Lonesome Optics, who
fabricated several optics for the instrument but died before its completion;
it is a joint project of the Hobby-Eberly Telescope partnership and the
Instituto de Astronom\'{\i}a de la Universidad Nacional Aut\'onoma de M\'exico.
This publication makes use of data products from
the USNOFS Image and Catalog Archive operated by the United States Naval
Observatory, Flagstaff Station, and the Two Micron All
Sky Survey, which is a joint project of the University of Massachusetts
and the Infrared Processing and Analysis Center/California Institute
of Technology, funded by the National Aeronautics and Space
Administration and the National Science Foundation.

\clearpage
\begin{deluxetable}{clllll}
\tablewidth{0pt}
\tablecaption{Observing Log\label{tab:log}}
\tablehead{
\colhead{} &
\colhead{} &
\colhead{} &
\colhead{} &
\colhead{$\lambda$} &
\colhead{} \\
\colhead{Night} &
\colhead{Date} &
\colhead{Telescope + Instrument} &
\colhead{Disperser} &
\colhead{($\mu$m)} &
\colhead{$\lambda/\Delta \lambda$} 
}
\startdata
1 & 2004 Nov 12 & IRTF + SpeX & prism & 0.8-2.5 & 100 \\
2 & 2004 Nov 13 & IRTF + SpeX & prism & 0.8-2.5 & 100 \\
3 & 2004 Dec 10 & MMT + Blue Channel & 600 grating  & 0.63-0.89 & 2900 \\
4 & 2004 Dec 11 & MMT + Blue Channel & 600 grating  & 0.63-0.89 & 2900 \\
5 & 2004 Dec 12 & MMT + Blue Channel & 600 grating  & 0.63-0.89 & 2900 \\
6 & 2005 Sep 25 & HET + LRS & G3 grism  & 0.63-0.91 & 1100 \\
7 & 2005 Sep 27 & HET + LRS & G3 grism  & 0.63-0.91 & 1100 \\
8 & 2005 Oct 21 & HET + LRS & G3 grism  & 0.63-0.91 & 1100 \\
9 & 2005 Dec 12 & IRTF + SpeX & prism & 0.8-2.5 & 100 \\
10 & 2005 Dec 13 & IRTF + SpeX & prism & 0.8-2.5 & 100 \\
\enddata
\end{deluxetable}

\begin{deluxetable}{llllll}
\tablewidth{0pt}
\tablecaption{Field Stars \label{tab:field}}
\tablehead{
\colhead{2MASS\tablenotemark{a}} &
\colhead{Spectral Type} &
\colhead{$J-H$\tablenotemark{a}} & \colhead{$H-K_s$\tablenotemark{a}}
& \colhead{$K_s$\tablenotemark{a}} &
\colhead{Night} 
}
\startdata
   J04095207+2821399 &     M4V &    0.79 &    0.53 &    13.86 &   4 \\
   J04111034+2830379 &     $<$M0 &    1.00 &    0.63 &    13.50 &   4 \\
   J04114008+2834024 &     $<$G0 &    0.72 &    0.35 &    11.99 &   4 \\
   J04122245+2827470 &     $<$K0 &    0.97 &    0.52 &    13.18 &   4 \\
   J04125785+2556088 &     M8V &    0.73 &    0.46 &    13.73 &   5 \\
   J04143306+3033411 &     M8-M9V(op,IR) &    0.68 &    0.47 &    13.80 &   5,10 \\
   J04151433+2840321 &     M1V &    0.86 &    0.46 &    13.61 &   5 \\
   J04153235+2908447 &     M4V &    1.01 &    0.54 &    13.67 &   4 \\
   J04172402+2837197 &   M3.5V &    1.01 &    0.65 &    13.34 &   5 \\
   J04172478+1634364 &   M8.5(op,IR)\tablenotemark{b} &    0.73 &    0.53 &    12.90 & 2,3 \\
   J04173180+2849444 &   M4.75V &    0.88 &    0.46 &    13.69 &   5 \\
   J04175041+2814403\tablenotemark{c} &    $<$M0 &    1.48 &    0.86 &    14.47 &   2 \\
   J04184416+2831533\tablenotemark{c} &    $<$M0 &    1.15 &    0.61 &    15.40 &   2 \\
   J04202573+2513013 &   M3.25V &    0.91 &    0.47 &    12.57 &   5 \\
   J04203904+2355502 &  L1(op,IR)\tablenotemark{b} &    0.85 &    0.59 &    13.50 &   5,10 \\
   J04214004+2853048 &     M3V &    1.19 &    0.70 &    12.55 &   4 \\
   J04220170+2653225 &     $<$K0 &    0.97 &    0.52 &    13.31 &   5 \\
   J04230586+2345204 &     $<$M0 &    1.11 &    0.66 &    13.31 &   5 \\
   J04251784+2641211 &   M3.5V &    0.89 &    0.59 &    13.80 &   5 \\
   J04260704+2430070 &     $<$K0 &    0.99 &    0.53 &    13.33 &   5 \\
   J04271561+1850364 &     $<$K0 &    0.84 &    0.42 &    12.35 &   5 \\
   J04272913+1854245 &     $<$K0 &    0.96 &    0.52 &    12.93 &   5 \\
   J04273708+2056389 &   M5.75V &    0.63 &    0.38 &    12.07 &   5 \\
   J04284746+1837356 &   M2.5V &    0.99 &    0.59 &    13.52 &   5 \\
   J04304017+2409526 &     $<$M0 &    0.89 &    0.50 &    13.35 &   4 \\
   J04310604+2409588 &     $<$M0 &    0.97 &    0.51 &    12.98 &   4 \\
   J04314195+2431268\tablenotemark{c} &    $<$M0 & 1.68 & 0.91 & 13.45 & 10 \\
   J04320000+2406343 &   M2.25V &   0.96 &    0.51 &    13.81 &   4 \\
   J04320865+2418583\tablenotemark{c} &    $<$M0 &    1.00 &    0.81 &    15.15 &   2 \\
   J04322947+2426174\tablenotemark{c} &      $<$M0 &  1.21 & 0.57 & 15.04 & 10 \\
   J04324813+2431063\tablenotemark{c} &   M8.25(op),M8(IR)\tablenotemark{b} &    0.60 &    0.66 &    14.74 & 2,4 \\
   J04330831+2413195\tablenotemark{c} &   L1-L3V &    0.71 &    0.63 &    15.40 &   2 \\
   J04331906+2343035 &   M3.75V &    0.70 &    0.47 &    13.32 &   4 \\
   J04332359+2650191 &     $<$M0 &    0.93 &    0.50 &    12.85 &   3 \\
   J04333296+2506587 &   M3.25V &    0.98 &    0.62 &    13.61 &   4 \\
   J04335918+2552238\tablenotemark{d} &    M4V &    0.73 &    0.74 &    14.94 &   2 \\
   J04341763+2251297 &     $<$M0 &    0.94 &    0.50 &    13.43 &   3 \\
   J04344356+1638484 &   M2.5V &    0.71 &    0.48 &    13.59 &   3 \\
   J04344701+1652593 &   M4.5V &    0.64 &    0.52 &    13.91 &   4 \\
   J04351796+2408105 &     $<$K0 &    0.88 &    0.45 &    11.66 &   4 \\
   J04352837+2410004\tablenotemark{d} &    $<$M0 &    0.90 &    0.74 &    15.23 &   2 \\
   J04404725+2501121 &     M4V &    1.08 &    0.63 &    13.35 &   5 \\
   J04440270+2515065 &   M3.25V &    0.83 &    0.49 &    13.58 &   5 \\
   J04470883+2921026 &   M7.5V &    0.67 &    0.46 &    13.60 &   5 \\
   J04474757+2819165 &   L2-L3V &    1.05 &    0.77 &    13.29 &   2 \\
   J04480068+1710115 &     $<$K0 &    0.79 &    0.38 &    11.90 &   4 \\
   J04481381+2559399 &     M1V &    0.85 &    0.47 &    13.52 &   4 \\
   J04482244+2051433 &     M6V &    0.62 &    0.41 &    12.28 &   4 \\
   J04483627+2541137 &     $<$K0 &    0.74 &    0.36 &    12.59 &   4 \\
   J04490477+2535229 &     M5V &    0.89 &    0.53 &    13.11 &   5 \\
   J04490506+1703513 &     M4V &    0.68 &    0.35 &    12.40 &   5 \\
   J04503796+2624466 &   M3.5V &    0.71 &    0.48 &    13.76 &   5 \\
   J04523836+2708423 &   M3.25V &    0.82 &    0.46 &    13.83 &   5 \\
   J04554336+2812523 &     M8V &    0.60 &    0.48 &    13.71 &   5 \\
   J04555897+2140007 &   M8-9V &    0.67 &    0.51 &    12.93 &   2 \\
\enddata
\tablenotetext{a}{2MASS Point Source Catalog.}
\tablenotetext{b}{Spectral features suggest a surface 
gravity intermediate between those of Taurus members and field dwarfs
(Figs.~\ref{fig:ir2} and \ref{fig:op4}).}
\tablenotetext{c}{Selected from photometry in \citet{luh00tau} 
(Figure~\ref{fig:4sh1}).}
\tablenotetext{d}{Selected from photometry in \citet{bri02}.}
\end{deluxetable}

\begin{deluxetable}{lllllllll}
\tabletypesize{\scriptsize}
%\rotate
\tablewidth{0pt}
\tablecaption{Members of Taurus\label{tab:mem}}
\tablehead{
\colhead{} &
\colhead{} &
\colhead{} &
\colhead{} &
\colhead{Membership} &
\colhead{} &
\colhead{} &
\colhead{} &
\colhead{} \\
\colhead{2MASS\tablenotemark{a}} &
\colhead{Other Names} &
\colhead{Spectral Type\tablenotemark{b}} &
\colhead{Ref} &
\colhead{Evidence\tablenotemark{c}} &
\colhead{$J-H$\tablenotemark{a}} & \colhead{$H-K_s$\tablenotemark{a}} & 
\colhead{$K_s$\tablenotemark{a}} &
\colhead{Night}}
%\tablecolumns{9}
\startdata
\sidehead{Previously Known Members}
   J04185813+2812234 & IRAS 04158+2805 &   M3,K7-M3,M6, & 1,2,3, & e,ex,$A_V$,NaK,H$_2$O &    1.43 &    1.17 &    11.18 & 1,3 \\
 & & M5.25(op),M6(IR)  & 4 & & & & & \\
   J04194148+2716070 & IRAS 04166+2706 &          $<$M0 &      4 &    e,ex &    1.30 &    0.79 &    12.62 &   5 \\
   J04221675+2654570 &    CFHT 21 &   M1.25,M1-M2 &  5,4 & e,$A_V$,NaK &    1.54 &    1.03 &     9.01 &   6 \\
   J04274538+2357243 &    CFHT 15 &        M8.25 &  5,4 &      NaK &    0.70 &    0.55 &    13.69 &   6 \\
   J04275730+2619183 & IRAS 04248+2612 &   M2,M5.5, & 1,3, & e,ex,NaK,$A_V$ &    1.44 &    0.77 &    11.03 & 1,3 \\
 & & M4.5(op),M4.75(IR) & 4 & & & & & \\
   J04292165+2701259 &    IRAS 04263+2654,CFHT 18 &     M6,M5.25 &  5,4 &   $A_V$,NaK &    1.30 &    0.77 &     8.72 &   6 \\
   J04295950+2433078 &    CFHT 20 &      M5.5,M5 &  5,4 & e,$A_V$,NaK &    1.15 &    0.73 &     9.81 &   8 \\
   J04302365+2359129 &    CFHT 16 &   M8.5,M8.25 &  5,4 &      NaK &    0.72 &    0.55 &    13.70 &   6 \\
   J04321786+2422149 &     CFHT 7 &   M6.5,M5.75 &  5,4 &      NaK &    0.75 &    0.41 &    10.38 &   8 \\
   J04325026+2422115 &   CFHT 5 & M7.5 & 5,4 & $A_V$,NaK,H$_2$O & 1.74 & 0.94 & 11.28 & 9 \\
   J04350850+2311398 &    CFHT 11 &     M6.75,M6 &  5,4 &      NaK &    0.59 &    0.35 &    11.59 &   7 \\
   J04400174+2556292 &    CFHT 17 &   M5.75,M5.5 &  5,4 &   $A_V$,NaK &    1.58 &    0.88 &    10.76 &   3 \\
   J04400800+2605253 & IRAS 04370+2559 &          $<$M0 &      4 &    e,ex &    2.16 &    1.38 &     8.87 &   3 \\
   J04422101+2520343 &     CIDA 7 &   M2-M3?,M4.75 &  6,4 & e,NaK,$A_V$ &    0.82 &    0.41 &    10.17 &   3 \\
   J04520668+3047175 & IRAS 04489+3042 &   M2,M4-M8, & 1,3, & e,ex,$A_V$ &    2.40 &    1.64 &    10.38 & 1,3 \\
 & & M3.5-M4.5(op),M3-M4(IR) & 4 & & & & & \\
\sidehead{New Members}
   J04080782+2807280 &         \nodata &        M3.75 &  4 &    Li,$A_V$ &    0.71 &    0.35 &    11.39 &   4 \\
   J04152409+2910434 &         \nodata &           M7 &  4 &    $A_V$,NaK &    0.80 &    0.53 &    12.36 &   5 \\
   J04161885+2752155 &         \nodata &        M6.25 &  4 &    $A_V$,NaK &    0.77 &    0.43 &    11.35 &   5 \\
   J04163911+2858491 &         \nodata &         M5.5 &  4 &    $A_V$,NaK &    0.88 &    0.56 &    11.28 &   5 \\
   J04214631+2659296 &    CFHT 10 &   M6.25,M5.75 & 5,4 &   $A_V$,NaK &    1.09 &    0.60 &    12.13 &   4 \\
   J04215450+2652315 &         \nodata &         M8.5(op),M8.75(IR) & 4 &     NaK?,H$_2$O &    1.04 &    0.60 &    13.90 &   4,10 \\
   J04221332+1934392 &         \nodata &           M8 &   4 &     NaK &    0.81 &    0.53 &    11.53 &   4 \\
   J04221644+2549118 &    CFHT 14 &        M7.75 &   5,4 &   NaK &   0.70 &    0.43 &    11.94 &   5 \\
   J04242646+2649503 &     CFHT 9 &   M6.25,M5.75 &  5,4 & Li,NaK &   0.69 &    0.43 &    11.76 &   5 \\
   J04263055+2443558 &         \nodata &        M8.75(op,IR) & 4 & e,NaK,H$_2$O &    0.72 &    0.54 &    13.40 & 2,3 \\
   J04290068+2755033 &         \nodata &        M8.25 &  4 &    e,NaK,H$_2$O &    0.69 &    0.47 &    12.85 &   5 \\
   J04295422+1754041 &         \nodata &           M4 &  4 &      Li,NaK &    0.94 &    0.69 &    11.02 &   5 \\
   J04311907+2335047\tablenotemark{d} &         \nodata &        M7.75 &  4 &        NaK &    0.79 &    0.52 &    12.20 &   4 \\
   J04312669+2703188 &    CFHT 13 &   M7.25,M7.5 & 5,4 &    NaK &    0.86 &    0.52 &    13.45 &   4 \\
   J04320329+2528078 &         \nodata &        M6.25 &  4 &      Li,NaK &    0.61 &    0.39 &    10.72 &   4 \\
   J04322329+2403013 &         \nodata &        M7.75 &  4 &     e,NaK &    0.64 &    0.36 &    11.33 &   4 \\
   J04330945+2246487 &    CFHT 12 &      M6.5,M6 & 5,4 &  e,$A_V$,NaK &    1.01 &    0.60 &    11.54 &   3 \\
   J04334291+2526470 &         \nodata &        M8.75(op,IR) & 4 &          NaK,H$_2$O  &    0.79 &    0.52 &    13.33 &   4,10 \\
   J04354526+2737130 &         \nodata &        M9.25 &      4 &      NaK &    0.77 &    0.53 &    13.71 &   5 \\
   J04361030+2159364 &         \nodata &         M8.5 &      4 &      NaK &    0.75 &    0.46 &    13.64 &   5 \\
   J04414489+2301513 &         \nodata &      M8.5(op),M8(IR)  & 4 & e,NaK,H$_2$O &    0.69 &    0.56 &    13.16 & 2,3 \\
   J04484189+1703374 &         \nodata &           M7 &      4 &      NaK &    0.60 &    0.44 &    12.49 &   4 \\
\enddata
\tablenotetext{a}{2MASS Point Source Catalog.}
\tablenotetext{b}{Uncertainties are $\pm0.25$ and $\pm0.5$ subclass for the 
optical and IR types from this work, respectively, unless noted otherwise.}
\tablenotetext{c}{Membership in Taurus is indicated by $A_V\gtrsim1$ and
a position above the main sequence for the distance of Taurus (``$A_V$"),
strong emission lines (``e"), Na~I and K~I strengths intermediate 
between those of dwarfs and giants (``NaK"), 
strong Li absorption (``Li"), IR excess emission (``ex"),
or the shape of the gravity-sensitive steam bands (``H$_2$O").}
\tablenotetext{d}{Also discovered by Slesnick, Carpenter, \& Hillenbrand 
(in preparation).}
\tablerefs{
(1) \citet{ken98};
(2) \citet{lr98};
(3) \citet{whi04};
(4) this work;
(5) \citet{gui06};
(6) \citet{bri99}.}
\end{deluxetable}
\clearpage

\begin{thebibliography}{}

\bibitem[Baraffe et al.(1998)]{bar98}
Baraffe, I., Chabrier, G., Allard, F., \& Hauschildt, P. H. 1998, \aap, 337, 403

\bibitem[Bate et al.(1998)]{bat98}
Bate, M. R., Clarke, C. J., \& McCaughrean, M. J. 1998, \mnras, 297, 1163

\bibitem[Bate et al.(2002)]{bat02}
Bate, M. R., Bonnell, I. A., \& Bromm, V. 2002, \mnras, 332, L65

\bibitem[Bate et al.(2003)]{bat03}
Bate, M. R., Bonnell, I. A., \& Bromm, V. 2003, \mnras, 339, 577



\bibitem[Boss(2001)]{bos01}
Boss, A. 2001, \apj, 551, L167


\bibitem[Brice\~no et al.(1997)]{bri97}
Brice\~{n}o, C., Hartmann, L., Stauffer, J., Gagne, M. Stern, R. A., \&
Caillault, J.-P. 1997, \aj, 113, 740

\bibitem[Brice\~no et al.(1998)]{bri98}
Brice\~{n}o, C., Hartmann, L., Stauffer, J., \& Mart{\'\i}n, E. L., 1998, \aj, 
115, 2074 

\bibitem[Brice\~no et al.(1999)]{bri99}
Brice\~{n}o, C., Calvet, N., Kenyon, S., \& Hartmann, L. 1999, \aj, 118, 1354

\bibitem[Brice\~no et al.(2002)]{bri02}
Brice\~{n}o, C., Luhman, K. L., Hartmann, L., Stauffer, J. R., \& Kirkpatrick, 
J. D. 2002, \apj, 580, 317





\bibitem[Chabrier et al.(2000)]{cha00}
Chabrier, G., Baraffe, I. Allard, F., \& Hauschildt, P. H. 2000, \apj, 542, 464



\bibitem[Cushing, Vacca, \& Rayner(2004)]{cus04}
Cushing, M. C., Vacca, W. D., \& Rayner, J. T. 2004, \pasp, 116, 362

\bibitem[Delgado-Donate, Clarke, \& Bate(2003)]{del03}
Delgado-Donate, E. J., Clarke, C. J., \& Bate, M. R. 2003, \mnras, 342, 926

\bibitem[Dobashi et al.(2005)]{dob05}
Dobashi, K., Uehara, H., Kandori, R., Sakurai, T., Kaiden, M., Umemoto, T.,
\& Sato, F. 2005, \pasj, 57, 1





\bibitem[Gomez et al.(1993)]{gom93}
Gomez, M., Hartmann, L., Kenyon, S. J., \& Hewett, R. 1993, \aj, 105, 1927


\bibitem[Guieu et al.(2006)]{gui06}
Guieu, S., Dougados, C., Monin, J.-L., Magnier, E. \& Mart{\'\i}n, E. L. 2006,
\aap, 446, 485 (G06)



\bibitem[Hartmann(2002)]{har02}
Hartmann, L. 2002, \apj, 578, 914



\bibitem[Herbig \& Bell(1988)]{hb88}
Herbig, G. H., \& Bell, K. R. 1988, Lick Obs.\ Bull.\ Ser., No.\ 1111





\bibitem[Kenyon \& Hartmann(1995)]{kh95}
Kenyon, S. J., \& Hartmann, L. 1995, \apjs, 101, 117

\bibitem[Kenyon et al.(1998)]{ken98}
Kenyon, S. J., Brown, D. I., Tout, C. A., \& Berlind, P. 1998, \aj, 115, 2491

\bibitem[Kirkpatrick et al.(1997)]{kir97}
Kirkpatrick, J. D., Henry, T. J., \& Irwin, M. J. 1997, \aj, 113, 1421

\bibitem[Kirkpatrick et al.(1999)]{kir99}
Kirkpatrick, J. D., et al. 1999, \apj, 519, 802

\bibitem[Kirkpatrick et al.(2006)]{kir06}
Kirkpatrick, J. D., et al. 2006, \apj, in press


\bibitem[Kroupa \& Bouvier(2003)]{kb03}
Kroupa, P., \& Bouvier, J. 2003, \mnras, 346, 369




\bibitem[Larson(1995)]{lar95}
Larson, R. B. 1995, \mnras, 272, 213


\bibitem[Lucas et al.(2001)]{luc01}
Lucas, P. W., Roche, P. F., Allard, F., \& Hauschildt, P. H. 2001,
\mnras, 326, 695

\bibitem[Luhman(1999)]{luh99}
Luhman, K. L. 1999, \apj, 525, 466

\bibitem[Luhman(2000)]{luh00tau} 
Luhman, K. L. 2000, \apj, 544, 1044

\bibitem[Luhman(2004a)]{luh04cha}
Luhman, K. L. 2004a, \apj, 602, 816

\bibitem[Luhman(2004b)]{luh04tau}
Luhman, K. L. 2004b, \apj, 617, 1216


\bibitem[Luhman \& Rieke(1998)]{lr98}
Luhman, K. L., \& Rieke, G. H. 1998, \apj, 497, 354

\bibitem[Luhman \& Rieke(1999)]{lr99}
Luhman, K. L., \& Rieke, G. H. 1999, \apj, 525, 440

\bibitem[Luhman et al.(2003a)]{luh03tau}
Luhman, K. L., Brice\~{n}o, C., Stauffer, J. R., Hartmann, L.,
Barrado y Navascu\'{e}s, D., \& Nelson, C. 2003a, \apj, 590, 348

\bibitem[Luhman et al.(2003b)]{luh03b}
Luhman, K. L., Stauffer, J. R., Muench, A. A., Rieke, G. H., Lada, E. A.,
Bouvier, J., \& Lada, C. J. 2003b, \apj, 593, 1093

\bibitem[Luhman et al.(2004)]{luh04ots}
Luhman, K. L., Peterson, D. E., \& Megeath, S. T. 2004, \apj, 617, 565

\bibitem[Luhman et al.(2005)]{luh05flam}
Luhman, K. L., Lada, E. A., Muench, A. A., \& Elston, R. J. 2005, \apj, 618, 810

\bibitem[Luhman, McLeod, \& Goldenson(2005)]{luh05wfpc}
Luhman, K. L., McLeod, K. K., \& Goldenson, N. 2005, \apj, 623, 1141

\bibitem[Mart{\'\i}n(2000)]{mar00}
Mart{\'\i}n, E. L. 2000, \aj, 120, 2114

\bibitem[Mart{\'\i}n et al.(2001)]{mar01}
Mart{\'\i}n, E. L., Dougados, C., Magnier, E., M\'{e}nard, F., Magazz\`{u}, A.,
Cuilandre, J.-C., \& Delfosse, X. 2001, \apj, 561, L195


\bibitem[McGovern et al.(2004)]{mc04}
McGovern, M. R., Kirkpatrick, J. D., McLean, I. S., Burgasser, A. J.,
Prato, L., \& Lowrance, P. J. 2004, \apj, 600, 1020

\bibitem[Monet et al.(2003)]{mon03}
Monet, D. G., et al. 2003, \aj, 125, 984







\bibitem[Nakajima et al.(1998)]{nak98}
Nakajima, Y., Tachihara, K., Hanawa, T., \& Nakano, M. 1998, \apj, 497, 721


\bibitem[Padoan \& Nordlund(2004)]{pn04}
Padoan, P., \& Nordlund, \AA. 2004, \apj, 617, 559



\bibitem[Rayner et al.(2003)]{ray03}
Rayner, J. T., et al. 2003, \pasp, 115, 362





\bibitem[Reipurth \& Clarke(2001)]{rc01}
Reipurth, B.~\& Clarke, C.\ 2001, \aj, 122, 432





\bibitem[Simon(1997)]{sim97}
Simon, M. 1997, \apj, 482, L81

\bibitem[Skrutskie et al.(2006)]{skr06}
Skrutskie, M. F., et al. 2006, \aj, 131, 1163



\bibitem[Strom \& Strom(1994)]{ss94}
Strom, K. M., \& Strom, S. E. 1994, \apj, 424, 237

\bibitem[Umbreit et al.(2005)]{umb05}
Umbreit, S., Burkert, A., Henning, T., Mikkola, S., \& Spurzem, R. 2005, \apj, 
623, 940

\bibitem[Vacca et al.(2003)]{vac03}
Vacca, W. D., Cushing, M. C., \& Rayner J. T., 2003, \pasp, 115, 389




\bibitem[White \& Hillenbrand(2004)]{whi04}
White, R. J., \& Hillenbrand, L. A. 2004, \apj, 616, 998



\end{thebibliography}
\end{document}